\documentclass[aps,prd,twocolumn,groupedaddress,showpacs]{revtex4-1}
\usepackage{amsmath,amssymb,IEEEtrantools}
\usepackage[dvips]{graphicx}
\usepackage[hidelinks,breaklinks=true]{hyperref}
\usepackage{cleveref}

\begin{document}

\title{Low-energy phenomenology of trinification: an effective left-right-symmetric model}

\author{Jamil Hetzel}
\email[]{hetzel@thphys.uni-heidelberg.de}
\author{Berthold Stech}
\email[]{b.stech@thphys.uni-Heidelberg.de}
\affiliation{Institut f\"ur Theoretische Physik, Universit\"at Heidelberg, D-69120 Heidelberg, Germany}

\date{\today}

\begin{abstract}
The trinification model is an interesting extension of the Standard Model (SM) based on the gauge group $SU(3)_C\times SU(3)_L\times SU(3)_R$.
We study its low-energy phenomenology by constructing a low-energy effective field theory, thereby reducing the number of particles and free parameters that need to be studied.
The resulting model predicts that several new scalar particles have masses in the $\mathcal{O}\left(100\text{ GeV}\right)$ range.
We study a few of the interesting phenomenological scenarios, such as the presence of a light fermiophobic scalar in addition to a SM-like Higgs, or a degenerate (twin) Higgs state at 126 GeV.
We point out regions of the parameter space that lead to measurable deviations from SM predictions of the Higgs couplings.
Hence the trinification model awaits crucial tests at the Large Hadron Collider in the coming years.
\end{abstract}


\maketitle

\newcommand{\ord}[1]{\mathcal{O}\left(#1\right)} 
\newcommand{\rep}[1]{\mathbf{#1}} 
\newcommand{\brep}[1]{\mathbf{\overline{#1}}} 
\newcommand{\vev}[1]{\langle{#1}\rangle{}} 
\newcommand{\trini}{{}SU(3)_C\times SU(3)_L\times SU(3)_R{}} 
\newcommand{\lag}{\mathcal{L}} 
\newcommand{\plushc}{+ \text{ h.c. }} 
\newcommand{\tr}[1]{\mathrm{Tr}\left\{#1\right\}} 
\newcommand{\LRgroup}{{}SU(3)_C\times SU(2)_L\times SU(2)_R\times U(1)_{B-L}{}} 
\newcommand{\SM}{{}SU(3)_C\times SU(2)_L\times U(1)_Y{}} 

\section{Introduction}
The discovery of the Higgs boson \cite{Aad:2012tfa,Chatrchyan:2012ufa} marks the establishment of the Standard Model (SM) of particle physics as the model that correctly describes physics at experimentally available energies to date.
All SM particles have been discovered, and the experimental data gathered at particle colliders match the predictions of the SM to good precision \cite{Baak:2013ppa}.
Yet, the SM is regarded to be an incomplete theory of nature: it lacks a dark matter candidate and is incompatible with the observation of non-zero neutrino masses.
Also, the fermion masses and mixings are free parameters that display hierarchical patterns, and parity violation is introduced by hand.
Therefore our quest towards a better theory of nature requires us to extend the SM.

Grand Unified Theories (GUTs) \cite{Georgi:1974sy,Fritzsch:1974nn,Gursey:1975ki,Achiman:1978vg,Shafi:1978gg} are interesting extensions of the SM in which the SM gauge group $\SM$ is embedded in a larger simple gauge group.
The exceptional group $E_6$ is an attractive example of a GUT group \cite{Gursey:1975ki,Achiman:1978vg,Shafi:1978gg}.
It is anomaly-free and left-right-symmetric (LR-symmetric), and as such it provides an explanation for parity violation in the SM by spontaneous symmetry breaking.
It appears in the compactification of string theories, which leads to either four-dimensional $E_6$ gauge symmetry or one of $E_6$'s maximal subgroups \cite{Candelas:1985en,Witten:1985xc}.
One of these maximal subgroups is the `trinification group' $G_{333} \equiv \trini$.
Models based on the trinification group have been studied in several contexts \cite{Cremmer:1979uq,Babu:1985gi,He:1986cs,Lazarides:1993uw,Lazarides:1994px,Demaria:2005gka,PhysRevD.83.093008,Sayre:2006ma}.

In this work, `trinification model' will refer to the setup described in refs.~\cite{Achiman:1978vg,Stech:2003sb,Stech:2008wd,Stech:2010gf,Stech:2012zr,Stech:2014tla}.
The setup described there is interesting for several reasons: fermion masses and mixings of the SM can be reproduced using only a few parameters, with a satisfactory fit for the solar neutrino mass difference and the neutrino mixing pattern.
Also, a Standard-Model-like Higgs with a mass close to 126 GeV appears in a large region of parameter space of the model.
Furthermore, it gives predictions for the matrix element of neutrinoless double-beta decay and the neutrino masses, which allow the model to be tested with low-energy experiments.
It also allows for various interesting phenomenological scenarios, such as the presence of a light fermiophobic Higgs in addition to the Standard-Model-like Higgs, or even a degenerate Higgs state at 126 GeV.

In order to compare the trinification model with experiment, a study of the low-energy phenomenology is necessary.
Due to the large number of scalars, a study of the full scalar mass matrix is challenging.
However, several of the scalar fields will obtain very large masses when the trinification symmetry is broken, and thus can be integrated out from the theory.
The result is an effective field theory with the LR-symmetric gauge group $\LRgroup$, and fewer scalar fields than in the trinification model.
This model has the same low-energy properties as the trinification model, but is easier to study.
We will refer to this model as the low-energy trinification (LET) model.

Neither the LET model nor the trinification model resolves the hierarchy problem.
This problem is hidden in the vacuum expectation values (vevs) used in the model, which are presently not understood.
In our treatment, all dimensionful parameters and masses are fully determined by these vevs multiplied by dimensionless coupling constants.
Since these vevs are momentum- and scale-independent (except for wave-function renormalization), their use as fixed parameters is justified.

Left-right symmetric models based on the gauge group $\LRgroup$ \cite{Pati:1974yy,Mohapatra:1974hk,Senjanovic:1975rk} have been studied extensively in the literature.
Moreover, these models have many features in common with the two-Higgs-doublet model (2HDM) \cite{Lee:1973iz}.
However, the LET model has properties that distinguish it from more general LR-symmetric models and the 2HDM, due to the trinification origin at high energy scales.
A LR-symmetric model in the context of the trinification model has not been studied before to the best of our knowledge.
Therefore the LET model merits a study.

In this work, we explore the low-energy phenomenology of the LET model.
We briefly describe the trinification model in \cref{s:trinificationModel}, and subsequently derive the properties of the LET model in \cref{s:LETmodel}.
In order to aid our study of the LET model, we first introduce the Single-Bidoublet (SB) model, a simplified form of the LET model that has fewer scalar fields.
The possible phenomenological scenarios for both the SB model and the LET model are worked out in \cref{s:trinificationPhenomenology}.
We show how these models may be distinguished experimentally from the SM by studying the modifications of the SM Higgs couplings in \cref{s:HCMs}.
It turns out that the LET model allows for interesting phenomenological scenarios, such as a very light fermiophobic scalar with a mass in the GeV range or a degenerate scalar state at 126 GeV.
We discuss these scenarios in more detail in \cref{s:VLF,s:twinHiggs}.
Our conclusions are presented in \cref{s:conclusions}.

\section{The trinification model}\label{s:trinificationModel}
The Higgs sector of the trinification model contains two complex scalar fields $H_1, H_2$ in the $(\rep{1},\brep{3},\rep{3})$ representation of $G_{333} \equiv \trini$.
We can use the $SU(3)_L\times SU(3)_R$ gauge symmetry to bring the vev of $H_1$ into diagonal form:
\begin{equation}
\vev{H_1} = \frac{1}{\sqrt2}\begin{pmatrix} v_1 & 0 & 0 \\ 0 & b_1 & 0 \\ 0 & 0 & M_1 \end{pmatrix}, \;
\vev{H_2} = \frac{1}{\sqrt2}\begin{pmatrix} v_2 & 0 & 0 \\ 0 & b_2 & b_3 \\ 0 & M & M_2 \end{pmatrix}. \label{eq:H1vev}
\end{equation}
Here we employed a matrix notation in which $SU(3)_L$ indices run vertically and $SU(3)_R$ indices run horizontally.
All vev parameters are taken to be real in order to avoid tree-level $CP$-violation.
The second field $H_2$ is necessary to break the left-right symmetry of $G_{333}$: it cannot be made diagonal once $\vev{H_1}$ is taken to be diagonal.
The off-diagonal parameters $M$, $b_3$ are taken to be unequal, and thus break the left-right symmetry.
We assume the presence of large hierarchies among the vev parameters.
The parameters $M_1,M_2 \sim 10^{13} \text{ GeV}$ are of the order of the scale where the Standard-Model gauge couplings $g_1$ and $g_2$ unify.
The off-diagonal vev $M$ is taken to be an intermediate scale of order $10^{10}$~GeV, but could also be as low as a few TeV.
The gauge couplings $g_{L,R}$ of $SU(3)_{L,R}$ are equal above this scale, whereas below $M$ the left-right symmetry is broken.
The other vev parameters contribute to the $W$-boson mass and are therefore much smaller than $M_1$, $M_2$, $M$.
As such, the former are constrained by the relation $v_1^2+v_2^2+b_1^2+b_2^2+b_3^2 = v^2 = (246\text{ GeV})^2$.
The used scalar potential is renormalizable and all its parameters are taken to be real in order to avoid tree-level $CP$-violation.

The fermions are grouped into the fundamental representation $\rep{27}$ of $E_6$.
They are two-component left-handed Weyl spinors with respect to the Lorentz group.
The fermion field decomposes into a lepton field $L$, a left-handed quark field~$Q_L$, and a right-handed quark field~$Q_R$, which are assigned to the representations of $\trini$ as follows:
\begin{equation}
L \sim (\rep{1},\brep{3},\rep{3}),\quad
Q_L \sim (\brep{3},\rep{3},\rep{1}),\quad
Q_R \sim (\rep{3},\rep{1},\brep{3}). \label{eq:trinifermionreps}
\end{equation}
In matrix notation, $L$ is a $3\times3$ matrix, $Q_L$ is a column vector, and $Q_R$ is a row vector:
\begin{align}
L = \left(\begin{array}{ccc} L^1_1 & E^- & e^- \\
E^+ & L^2_2 & \nu \\
e^+ & \hat\nu & L^3_3 \end{array}\right),\;
Q_L^b =& \left(\begin{array}{c} u^b \\ d^b \\ D^b \end{array}\right), \notag\\
Q_R^b =& \left(\begin{array}{ccc} \hat{u}^b & \hat{d}^b & \hat{D}^b \end{array}\right). \label{eq:trinifermions}
\end{align}
Here $b=1,2,3$ is a color index.
The components $u$, $d$ are the left-handed up-type and down-type quarks from the SM, whereas $D$ is a new quark with electromagnetic charge $-\frac13$.
The components $\hat{u}$, $\hat{d}$, $\hat{D}$ are their respective right-handed counterparts.
The lepton field contains the charged leptons $e^\pm$ and the left-handed neutrino $\nu$.
It contains several new states: a right-handed neutrino $\hat\nu$; three neutral states $L^1_1$, $L^2_2$, $L^3_3$; and a pair of charged leptons $E^\pm$.
A generation index $\alpha=1,2,3$ on the fermion fields in \cref{eq:trinifermions} has been suppressed.

The Higgs fields $H_1$, $H_2$ cannot both couple to fermions, since this would lead to flavor-changing neutral current (FCNC) processes, which are severly restricted by experiment.
In order to suppress tree-level FCNC interactions, the existence of a $Z_2$-symmetry is assumed under which $H_1$ ($H_2$) is even (odd).
The fermions are even under this symmetry as well, which implies that $H_2$ does not couple to fermions.
The Yukawa couplings are of the form
\begin{align}
\lag_Y =& -g_tG_{\alpha\beta} \left( Q_R^\alpha H_1^TQ_L^\beta + \frac12\epsilon^{ijk}\epsilon_{lmn}L^i_lL^j_m(H_1)^k_n \right) \notag\\
&- A_{\alpha\beta}\left( Q_R^\alpha H_{Aq}^T Q_L^\beta + \epsilon^{ijk}L^i_lL^j_m(H_{Al})^k_{\{lm\}} \right) \notag\\
&-\frac{1}{M_N} (G^2)_{\alpha\beta} \tr{L^\alpha H_1^\dagger} \tr{H_2^\dagger L^\beta} \plushc \label{eq:trinificationYukawa}
\end{align}
The first line is a Yukawa interaction built from the fields we have already introduced: the parameter $g_t$ is a dimensionless coupling, $G_{\alpha\beta}$ is a symmetric $3\times3$ generation matrix, and $\epsilon$ is a totally antisymmetric symbol with $\epsilon^{123} = \epsilon_{123} = +1$.
This interaction is sufficient to reproduce the up-quark masses by choosing a generation basis in which $g_tG_{\alpha\beta}$ is diagonal and fitting its diagonal components to the up-quark masses \cite{Stech:2014tla}.
The vev $M_1$ in \cref{eq:H1vev} gives large masses to $D$, $E^\pm$.
The second line in \cref{eq:trinificationYukawa} contains interactions with new scalar fields $H_{Aq} \sim (\rep{1},\brep{3},\rep{3})$ and $H_{Al} \sim (\rep{1},\brep{3},\brep{6})$, coupling to the fermions with a Hermitian antisymmetric matrix $A_{\alpha\beta}$.\footnote{The matrices $G_{\alpha\beta}$ and $A_{\alpha\beta}$ can be viewed respectively as the real and imaginary components of the vev of a flavon field (see ref.~\cite{Stech:2008wd}). In this picture, the Yukawa interactions are effective interactions arising from dimension-five operators, which in turn arise from interactions with gauge-singlet fermions. However, the components of these matrices are simply considered as free parameters of the trinification model.}
These terms come from the couplings of a scalar field $H_A$ in the antisymmetric $\rep{351_A}$ representation of $E_6$.
This interaction is necessary to describe the masses and mixings of the down quarks and charged leptons correctly: the Standard-Model down quarks and charged leptons are mixed with their heavy partners via the seesaw mechanism.
A good fit for the masses and mixings of the Standard-Model fermions is obtained using only very few extra parameters \cite{Stech:2014tla}.
It is assumed that the fields $H_{Aq}$, $H_{Al}$ have negligible mixing with $H_1$, $H_2$ in order to simplify the analysis of the scalar spectrum.

At this stage, neutrinos are still Dirac particles with masses comparable to the other fermion masses.
The third line in \cref{eq:trinificationYukawa} is necessary to obtain neutrino masses in accordance with experiment.
This line contains an effective dimension-five Yukawa interaction that could originate from the exchange of a new heavy Dirac fermion that is a trinification singlet \cite{Stech:2008wd}.
It violates the $Z_2$-symmetry and mixes the neutrinos $\nu$, $\hat\nu$ with the other neutral leptons $L^1_1$, $L^2_2$, $L^3_3$, giving rise to a generalised seesaw mechanism.
The light-neutrino mass matrix introduces two additional parameters, which can be fixed by the experimentally observed atmospheric mass-squared difference and the lightest neutrino mass \cite{Stech:2014tla}.

\section{An effective trinification model: the LET model}\label{s:LETmodel}
We now consider the effective trinification model obtained after integrating out the Higgs fields that are made heavy by the large vevs $M_1$ and $M_2$.
We expect the Higgs fields that are right-handed singlets with respect to the SM to obtain large masses.
These are the fields with the $SU(3)_R$ index (3).
On the other hand, the fields (3,1) and (3,2) of $H_2$ that are left-handed singlets (with respect to the SM) are kept.
These Higgs fields are necessary to describe the breaking of the left-right symmetry.
The corresponding scale $M$ is certainly larger than the weak scale, but may be much lower than $M_1$ and $M_2$.

The $2\times2$ blocks in the upper left corners of the fields $H_1$, $H_2$ transform as bidoublets $\Phi_1, \Phi_2 \sim (\rep{1},\brep{2},\rep{2},0)$ under $\LRgroup$.
The $(3,1)$ and $(3,2)$ components of $H_2$ transform as a right-handed doublet $\Phi_R \sim (\rep{1},\rep{1},\rep{2},1)$:
\begin{align}
\Phi_i = \begin{pmatrix} \Phi_{i,11}^0 & \Phi_{i,21}^- \\
\Phi_{i,12}^+ & \Phi_{i,22}^0 \end{pmatrix}
\leftrightarrow& \begin{pmatrix} (H_i)^1_1 & (H_i)^1_2 & 0 \\
(H_i)^2_1 & (H_i)^2_2 & 0 \\
0 & 0 & 0 \end{pmatrix}, \notag\\
\Phi_R = \begin{pmatrix} \Phi_R^+ & \Phi_R^0 \end{pmatrix}
\leftrightarrow& \begin{pmatrix} 0 & 0 & 0 \\
0 & 0 & 0 \\
(H_2)^3_1 & (H_2)^3_2 & 0 \end{pmatrix}.
\end{align}
These fields obtain the following vevs:
\begin{equation}
\vev{\Phi_i} = \frac{1}{\sqrt2}\begin{pmatrix} v_i & 0 \\
0 & b_i \end{pmatrix},\quad
\vev{\Phi_R} = \frac{1}{\sqrt2}\begin{pmatrix} 0 & M \end{pmatrix}. \label{eq:Phivev}
\end{equation}
The fields $\Phi_1$, $\Phi_2$, $\Phi_R$, and their vevs are sufficient to describe the symmetry breaking from $\LRgroup$ to electromagnetism via the SM.
Note that a right-handed doublet like $\Phi_R$ resides in $H_1$ as well.
In principle, $\Phi_R$ may be a combination of both.
Note that the $(3,3)$ components of $H_1$ and $H_2$ are total gauge singlets.
A combination of them, if sufficiently light, could be a dark matter candidate.
Also note that $\Phi_2$ and $\Phi_R$ do not couple to fermions.

Besides the eight gluon fields of $SU(3)_C$, the gauge-boson sector consists of seven vector gauge bosons.
Their masses and mixings are given in \cref{a:gaugebosons}.
The vevs of $\Phi_{1,2}$ contribute to the $W$ mass and are therefore constrained by the relation $v_1^2+b_1^2+v_2^2+b_2^2 = v^2 = (246\text{ GeV})^2$.
It is convenient to reparameterize the vev parameters as
\begin{align}
v_1 =& v\cos\alpha\cos\beta_1,\qquad v_2 = v\sin\alpha\cos\beta_2, \notag\\
b_1 =& v\cos\alpha\sin\beta_1,\qquad b_2 = v\sin\alpha\sin\beta_2. \label{eq:completeLETvevRedefinition}
\end{align}
The parameter $M$ is a mass scale above the electroweak scale.
Because of this hierarchy, it will often be convenient to use the small parameter $\xi \equiv v/M$.

We build the scalar potential of the LET model from all possible gauge-invariant renormalizable operators consisting of $\Phi_1$, $\Phi_2$, $\Phi_R$, leaving out those that could not have arisen from the trinification model.
This means that we do not include operators involving charge conjugates: since the $\rep{3}$ and $\brep{3}$ representations of $SU(3)$ are inequivalent, such operators have no possible origin in the trinification model.
The resulting potential is
\begin{align}
V =& V_1(\Phi_1,\Phi_R) + V_2(\Phi_1,\Phi_2,\Phi_R), \notag\\
V_1 =& \frac{\lambda_1}{2}\tr{\Phi_1^\dagger\Phi_1}^2 + \frac{\lambda_2}{2}\tr{\Phi_1^\dagger\Phi_1\Phi_1^\dagger\Phi_1} + \frac{\lambda_3}{2}\big(\Phi_R\Phi_R^\dagger\big)^2 \notag\\
&+ \lambda_4\tr{\Phi_1^\dagger\Phi_1}(\Phi_R\Phi_R^\dagger) + \lambda_{5}\Phi_R\Phi_1^\dagger\Phi_1\Phi_R^\dagger \notag\\
&+ \mu^2_{11}\tr{\Phi_1^\dagger\Phi_1} + \mu^2_R\Phi_R\Phi_R^\dagger + \left(\mu^2_1\det\Phi_1 \plushc \right), \notag\\
V_2 =& \frac{\widetilde\lambda_1}{2}\tr{\Phi_2^\dagger\Phi_2}^2 + \frac{\widetilde\lambda_2}{2}\tr{\Phi_2^\dagger\Phi_2\Phi_2^\dagger\Phi_2} \notag\\
&+ \widetilde\lambda_3\tr{\Phi_2^\dagger\Phi_2}(\Phi_R\Phi_R^\dagger) + \widetilde\lambda_4\Phi_R\Phi_2^\dagger\Phi_2\Phi_R^\dagger \notag\\
&+ \widetilde\lambda_5\tr{\Phi_1^\dagger\Phi_1}\tr{\Phi_2^\dagger\Phi_2} + \widetilde\lambda_6\left|\tr{\Phi_1^\dagger\Phi_2}\right|^2 \notag\\
&+ \frac{\widetilde\lambda_7}{2}\left( \tr{\Phi_1^\dagger\Phi_2}^2 \plushc\right) + \widetilde\lambda_8\tr{\Phi_1^\dagger\Phi_1\Phi_2^\dagger\Phi_2} \notag\\
&+ \widetilde\lambda_9\tr{\Phi_1^\dagger\Phi_2\Phi_2^\dagger\Phi_1} + \frac{\widetilde\lambda_{10}}{2}\left( \tr{\Phi_1^\dagger\Phi_2\Phi_1^\dagger\Phi_2} \plushc\right) \notag\\
&+ \mu^2_{22}\tr{\Phi_2^\dagger\Phi_2} + \left(\mu^2_2\det\Phi_2 \plushc \right). \label{eq:completeLETpotential}
\end{align}
For later convenience, we have split the potential into the parts $V_1$ and $V_2$ that respectively do and do not depend on $\Phi_2$.
As mentioned in the introduction, the five dimensionful parameters $\mu_{11}^2$, $\mu_{22}^2$, $\mu_R^2$, $\mu_1^2$, $\mu_2^2$ are no free parameters.
They are determined in terms of the vev parameters $v_1$, $b_1$, $v_2$, $b_2$, $M$ and the dimensionless parameters $\lambda_i$, $\widetilde{\lambda}_j$ by the requirement that $V$ has an extremum at the appropriate place.

In the trinification model, the fermions obtain their masses from Yukawa interactions with $H_1$.
In order to combine the leptons and $\Phi_1$ into a gauge singlet, we need the antisymmetric tensor $i\sigma_2$, which can be absorbed into a redefinition of the lepton fields.
Absorbing a minus sign into the phase of the $e^\pm$ fields, the fermionic field content of the LET model becomes
\begin{multline}
Q_L \equiv \begin{pmatrix} u \\ d \end{pmatrix} \sim (\brep{3},\rep{2},\rep{1}, \tfrac13),\; 
Q_R \equiv \begin{pmatrix} \hat{u} & \hat{d} \end{pmatrix} \sim (\rep{3},\rep{1}, \brep{2}, -\tfrac13), \\
L^- \equiv \begin{pmatrix} \nu \\ e^- \end{pmatrix} \sim (\rep{1},\rep{2}, \rep{1}, -1),\;
L^+ \equiv \begin{pmatrix} \hat\nu & e^+ \end{pmatrix} \sim (\rep{1},\rep{1}, \brep{2}, 1). \label{eq:fermionreps}
\end{multline}
After integrating out the heavy fields, the first line of the Yukawa Lagrangian in \cref{eq:trinificationYukawa} becomes
\begin{equation}
\lag_Y = -G_{\alpha\beta}\left( Q_R^\alpha \Phi_1^TQ_L^\beta + L^{+\alpha}\Phi_1^T L^{-\beta} \right) \plushc \label{eq:yukawalag}
\end{equation}
This term is insufficient to describe all fermion masses correctly: down-quark masses as well as the charged-lepton and neutrino masses would be proportional to each other and the CKM matrix would be a unit matrix at this point.
In order to describe the fermion masses and mixings correctly, interactions with the additional fields $H_{Aq}$, $H_{Al}$ need to be included.
Here we restrict ourselves to the single Yukawa term in \cref{eq:yukawalag} and fit the free parameters to the top- and bottom-quark masses: these fermions are the most relevant to compare our analysis to experimental searches for new physics.
We assume that flavor physics for the lower fermion masses does not influence the spectrum of the scalar particles.
Fitting the free parameters to the top- and bottom-quark masses, we find $v = 246$ GeV, $\tan\beta_1 = m_b/m_t \Rightarrow \beta_1 = 0.0166$.
Since $\Phi_2$ does not contribute to the fermion masses, we have no such restrictions on $\beta_2$ and $\alpha$.

\subsection{The Single-Bidoublet model}
The scalar potential of the LET model in \cref{eq:completeLETpotential} contains 15 free coupling constants, which are difficult to deal with.
As an intermediate step towards an understanding of the LET model, we first discuss a model from which $\Phi_2$ has been omitted; we will refer to this setup as the Single-Bidoublet (SB) model.
This model corresponds to the LET-model limit $\alpha = 0$, $\mu_{22}^2 \rightarrow \infty$ ($\alpha=0$ implies $v_2 = b_2 = 0$, for which $\mu_{22}^2$ is no longer constrained by the location of the minimum of the scalar potential).
Note however that the SB model is not an appropriate effective field theory of the trinification model, since the vev parameters of $\Phi_2$ are of order $v$ for the general case $\alpha \neq 0$.
Rather, we use it as a toy model that helps us study the phenomenology of the LET model.

The most general scalar potential for the SB model is given by $V_1(\Phi_1,\Phi_R)$ in \cref{eq:completeLETpotential}.
The minimalisation of this potential at the vev in \cref{eq:Phivev} fixes the dimensionful parameters $\mu^2_{11}$, $\mu^2_R$, $\mu^2_1$ in terms of the $\lambda_i$ and the vev parameters $v_1$, $b_1$, $M$.
The vev parameters $v_1$, $b_1$ can be reparameterized in terms of $v$, $\beta_1$ by \cref{eq:completeLETvevRedefinition}, with $\alpha=0$.

The scalar fields $\Phi_1$, $\Phi_R$ contain twelve real scalar components in total.
After spontaneous symmetry breaking, six of them become massless Goldstone bosons that give mass to the six massive vector gauge bosons.
The remaining components mix to form six massive scalars: three $CP$-even scalars $h^0$, $H_1^0$, and $H_2^0$, one $CP$-odd scalar $A^0$, and a pair of charged scalars $H^\pm$.
Their definitions in terms of gauge eigenstates as well as their masses are given in \cref{a:scalarSpectrum}.
We identify $h^0$ with the Standard-Model-like Higgs particle that has been observed at the LHC \cite{Aad:2012tfa,Chatrchyan:2012ufa}, since it is the only scalar that naturally has a mass at the electroweak scale.
The other scalars have masses of order $M$ unless we give $\ord{\xi^2} = \ord{v^2/M^2}$ values to some of the dimensionless scalar-potential parameters.

Note that the SB model resembles the two-Higgs-doublet model (2HDM) \cite{Lee:1973iz} (see \cite{Branco:2011iw} for a recent review).
The 2HDM is an extension of the SM in which the scalar sector contains an additional $SU(2)_L$ doublet.
It has been studied extensively since it provides a low-energy description of various models such as supersymmetry (see e.g.~\cite{Martin:1997ns} for a review), composite Higgs models \cite{Kaplan:1983sm}, and little Higgs models \cite{ArkaniHamed:2001nc}.

It is easy to see why the SB model resembles the 2HDM at the Lagrangian level.
To this end, we write $\Phi_1 = (i\sigma_2\phi_1, \phi_2^*)$ and $\Phi_R = (\phi_+, \phi_0)$, where $\phi_{1,2}$ are $SU(2)_L$ doublets and $\phi_{+,0}$ are $SU(2)_L$ singlets.
Since $\Phi_R$ has a much larger vev $M$ than the vev components $v_1$, $b_1$ of $\Phi_1$, the mixing among $\Phi_1$, $\Phi_R$ will be of order $\xi \ll 1$.
If we set $\Phi_R=0$ in the scalar potential $V_1$ in \cref{eq:completeLETpotential}, we can rewrite the entire scalar potential in terms of $\phi_1$, $\phi_2$ only.
The result is a 2HDM potential (see e.g.\ eq.~(98) in ref.~\cite{Branco:2011iw}) with the following constraints on the scalar parameters:
\begin{align}
\lambda^\text{2HDM}_{1,2,3} =& \lambda_1+\lambda_2,\quad \lambda^\text{2HDM}_4 = -\lambda_2,\quad \lambda^\text{2HDM}_{5,6,7} = 0, \nonumber\\
m_{11}^2 = m_{22}^2 =& \mu_{11}^2,\qquad m_{12}^2 = -\mu_1^2.
\end{align}
We can rewrite the Yukawa sector of the LET model in terms of $\phi_1$, $\phi_2$ as well.
In the Lagrangian in \cref{eq:yukawalag}, the up-type fermions couple only to $\phi_1$ whereas the down-type fermions couple only to $\phi_2$.
Hence the SB model resembles a constrained type-II 2HDM setup.
It differs from the 2HDM due to the presence of two additional heavy $SU(2)_L$ singlets.
The neutral one gives rise to an additional physical particle $H_2^0$ that is fermiophobic.
Its mass can be tuned independently from the masses of the 2HDM-like scalars $H_1^0$, $A^0$, $H^\pm$ (see \cref{eq:scalarmasses}).
This can result in phenomenological scenarios that cannot appear in the 2HDM.
Moreover, the vev ratio $\tan\beta \equiv \vev{\phi_2}/\vev{\phi_1}$ is a free parameter in the 2HDM, whereas $\tan\beta_1 = m_b/m_t$ is fixed in the LET model.
Likewise, the mixing angle $\alpha^\text{2HDM}$ of the $CP$-even scalars in the 2HDM can be taken to be a free parameter: the $\lambda_i^\text{2HDM}$ are usually rewritten in terms of the scalar masses and $\alpha^\text{2HDM}$.
However, the three mixing angles of the $CP$-even scalars in the SB model cannot be treated as free parameters: they are approximately fixed by the value of $\beta_1$ unless at least one of the new scalars becomes light.

\subsection{LET-model scalar spectrum}
Now let us consider the scalar sector of the LET model.
Compared to the SB model, it contains an additional bidoublet $\Phi_2$ with eight real scalar components.
This makes 14 physical scalars in total: five $CP$-even states $h^0$, $H^0_1$, $H^0_2$, $H^0_3$, $H^0_4$; three $CP$-odd states $A^0_1$, $A^0_2$, $A^0_3$; and three pairs of charged states $H^\pm_1$, $H^\pm_2$, $H^\pm_3$.
Their definitions and masses are given in \cref{a:scalarSpectrum}.
The $CP$-even states $h^0$, $H^0_1$ have masses of order $v$, whereas $H^0_2$, $H^0_3$, $H^0_4$ have masses of order $M$.
This is not surprising: if we decouple the bidoublet $\Phi_2$ from the model, we get one light state $h^0$ and two heavy states.
Since $\Phi_1$ and $\Phi_2$ are copies of the same representation, we expect that $\Phi_2$ adds one light and one heavy scalar to the spectrum as well.
The $CP$-odd state $A^0_1$ is light, whereas $A^0_2$, $A^0_3$ are heavy.
Again, this is not surprising.
In the SB model, the $CP$-odd components of the bidoublet $\Phi_1$ give rise to one Goldstone and one heavy state.
Thus we would expect $\Phi_2$ to contribute one heavy state as well.
Since there are no more would-be Goldstones, the other $CP$-odd component of $\Phi_2$ becomes a massive state with a mass of order $v$.
Similarly, $H^\pm_1$ is light whereas $H^\pm_2$, $H^\pm_3$ have masses of order $M$.

\section{Trinification phenomenology}\label{s:trinificationPhenomenology}
\subsection{The SB model}
Now we turn to the phenomenological scenarios that are allowed by the scalar sector of the trinification model.
As a first step towards understanding the trinification phenomenology, we consider the phenomenological scenarios that are possible in the SB model.
The free parameter space of the SB model is spanned by $M$ and the five scalar parameters $\lambda_i$.
A full analysis of this parameter space and the possible signatures is beyond the scope of this work.
Instead, we define a set of benchmark points that lead to distinct phenomenological features.
To get a feel for the possibilities, consider the scalar masses given in \cref{eq:scalarmasses}.
The mass of $h^0$ can be adjusted by changing the values of $\lambda_1$, $\lambda_2$, $\lambda_3$, $\lambda_4$.
We tune these parameters such that $m_{h^0} = 126$ GeV for each benchmark.
The leading contributions to $m_{H^0_1}$, $m_{A^0}$, $m_{H^\pm}$ are all given by $\sqrt{\lambda_5}M$, so we expect them to have similar masses, with $\ord{v}$ mass splittings.
On the other hand, $m_{H^0_2}$ is proportional to $\sqrt{\lambda_3}M$, which can be tuned independently of the other scalar masses.
Thus we expect the SB model to allow for different mass hierarchies or compressed spectra, depending on the magnitudes of $\lambda_3$, $\lambda_5$.

If any of the parameters $\lambda_3$, $\lambda_5$ have $\ord{\xi^2}$ values, some of the new scalars may obtain $\ord{v}$ masses.
Thus a number of phenomenologically different scenarios are possible.
If $\lambda_3$, $\lambda_5$ are not too small, all new scalars obtain $\ord{M}$ masses, beyond experimental reach.
We denote these benchmarks as having a Single Large Hierarchy (SLH); an additional hierarchy between $H^0_1$, $A^0$, $H^\pm$ on one side and $H_2^0$ on the other side is possible, but of little phenomenological interest.
For $\lambda_5 \sim \ord{\xi^2}$ but not too small $\lambda_3$, the 2HDM-like scalars $H_1^0$, $A^0$, $H^\pm$ all have $\ord{v}$ masses and could be observed at the LHC.
We choose two parameter sets such that these particles have masses in the $\ord{100\text{ GeV}}$ ballpark, and denote these benchmarks as 2HDM-1 and 2HDM-2 since they have a 2HDM-like spectrum at low energies.
For $\lambda_3 \sim \ord{\xi^2}$ but sizable $\lambda_5$, the fermiophobic state $H_2^0$ lies within experimental reach\footnote{Note that other models allow for the existence of a light fermiophobic Higgs as well, such as the type-I 2HDM and models with $SU(2)_L$-triplet Higgs fields \cite{Akeroyd:1995hg,Gunion:1989ci,Bamert:1993ah}. However, in those scenarios the light Higgs $h^0$ is fermiophobic, whereas we consider benchmark scenarios with a fermiophobic light Higgs \emph{in addition to} the SM-like Higgs.}.
We choose two parameter sets such that $m_{H^0_2}$ lies in the $\ord{100\text{ GeV}}$ range and denote these benchmarks as Light Fermiophobic (LF).
Note that for small $\lambda_3$, the parameters $\lambda_4$, $\lambda_5$ need to be chosen sufficiently small to ensure that $m_{h^0}^2$ remains positive (see \cref{eq:scalarmasses}).
This means that the 2HDM-like scalars have masses well below $M$.
A combination of the 2HDM-like and LF scenarios is possible as well: if both $\lambda_3$ and $\lambda_5$ are sufficiently small, all new scalars could have masses within experimental reach.
Again, we choose two parameter sets and refer to these benchmarks as Compressed for having a spectrum compressed around the electroweak scale.

We also consider two special cases of the LF scenario.
For small enough $\lambda_3$, $H^0_2$ can have a mass in the $\ord{1\text{ GeV}}$ ballpark.
Such a state would decay into pairs of photons only, since $H_2^0$ is fermiophobic.
If the signal strength for its decay is low enough, it could have escaped detection so far; we discuss the relevant experimental constraints in more detail in \cref{s:VLF}.
We refer to this scenario as having a Very Light Fermiophobic (VLF) Higgs.
The second special case is when $h^0$ and $H^0_2$ are approximately degenerate.
If their mass difference is less than their widths, both states would contribute to the signal strength used for the Higgs discovery, leading to a `twin Higgs'\footnote{Our use of the term `twin Higgs' is not to be confused with `twin Higgs models' in the literature. In those models, each Standard-Model particle has a corresponding particle that transforms under a mirror copy of the SM gauge group (see e.g.~refs.~\cite{Chacko:2005pe,Chacko:2005vw}). The copies are related by a $Z_2$ symmetry called `twin parity', and the twin Higgs is the partner of the Standard-Model Higgs.} \cite{Stech:2013pda,Heikinheimo:2013cua}.
This might result in deviations of the measured Higgs couplings from their Standard-Model values.
We tweak the parameters such that $m_{h^0,H_2^0} = 126\text{ GeV}$ and denote the corresponding benchmarks as Twin-1 and Twin-2.

For each benchmark point, we use the values $v=246$ GeV, $\beta_1=0.0166$ as well as the experimental values $\sin^2\theta_W=0.23126$, $g_L=0.65170$, and our best fit $\theta_W^\prime = 0.62$ as in \cref{a:gaugebosons}.
The other gauge couplings are fixed by the identities $g_R = g_L\tan\theta_W/\sin\theta_W^\prime$, $2g' = g_L\tan\theta_W/\cos\theta_W^\prime$.
As for the scale $M$, we consider both a high scale $M=10^{10}$ GeV well outside experimental reach and a lower scale $M=10^4$ GeV just beyond LHC reach.
We ensure that the constraints for vacuum stability and S-matrix unitarity are satisfied.
Multi-Higgs potentials in general can have several minima.
Determining the global minimum of the potential is already challenging in the 2HDM \cite{Barroso:2007rr}.
We could not check that our minimum is the global one.
The corresponding parameter values are given in \cref{t:benchmarkpoints}.
Note that except for the SLH scenario, all benchmarks contain very small values for $\lambda_3$, $\lambda_4$ and/or $\lambda_5$: these parameters need to have $\ord{v^2/M^2}$ values to compensate for the large vev of the corresponding scalar invariants.
This makes these scenarios unnatural.

To calculate the scalar masses, we do not use the approximations in \cref{eq:scalarmasses}, since subleading terms in $\xi$ may become large for small $\lambda_3$ and/or $\lambda_5$.
Instead, we evaluate the mass matrix numerically in \emph{Mathematica} \cite{Mathematica10} and then extract the masses and mass eigenstates.
The corresponding particle masses are given in \cref{t:benchmarkMasses}.
The scalar mass eigenstates are almost equal to the gauge eigenstates in most benchmarks: mixings are mostly below the percent level.
In the Compressed-2 benchmark, there is a 2\% mixing of $h^0_{1,11}$ with $h^0_R$.
Only Twin-2 gives large scalar mixing: in terms of squares of amplitudes, the SM-like Higgs $h^0$ is 62\% $h^0_{1,11}$ and 38\% $h^0_R$, whereas the fermiophobic Higgs $H^0_2$ is 38\% $h^0_{1,11}$ and 62\% $h^0_R$; mixing with $h^0_{1,22}$ is negligible.

\begin{table}
\begin{center}
\begin{tabular}{|l|l|c|c|c|c|c|}
\hline
Benchmark &	$M$	&	$\lambda_1$	&	$\lambda_2$	&	$\lambda_3$	&	$\lambda_4$	&	$\lambda_5$	\\\hline\hline
SLH-1	&	$10^{10}$	&	0.24	&	0.24	&	0.47	&	0.32	&	0.2	\\
SLH-2	&	$10^4$	&	0.24	&	0.24	&	0.47	&	0.32	&	0.2	\\\hline
2HDM-1	&	$10^{10}$	&	0.41	&	0.4	&	0.44	&	0.49	&	$5\cdot10^{-15}$	\\
2HDM-2	&	$10^4$	&	0.41	&	0.4	&	0.44	&	0.49	&	$5\cdot10^{-3}$	\\\hline
LF-1	&	$10^{10}$	&	0.133	&	0.13	&	$2\cdot10^{-15}$	&	$1\cdot10^{-12}$	&	$3\cdot10^{-7}$	\\
LF-2	&	$10^4$	&	0.14	&	0.14	&	$2\cdot10^{-3}$	&	$5.5\cdot10^{-3}$	&	0.6	\\\hline
Compressed-1	&	$10^{10}$	&	0.133	&	0.13	&	$1.1\cdot10^{-15}$	&	$1\cdot10^{-12}$	&	$5\cdot10^{-15}$	\\
Compressed-2	&	$10^4$	&	0.15	&	0.14	&	$1.1\cdot10^{-3}$	&	$5.2\cdot10^{-3}$	&	$5\cdot10^{-3}$	\\\hline
VLF-1	&	$10^{10}$	&	0.133	&	0.13	&	$1\cdot10^{-20}$	&	$1\cdot10^{-14}$	&	$2\cdot10^{-8}$	\\
VLF-2	&	$10^4$	&	0.13	&	0.13	&	$4.5\cdot10^{-7}$	&	$4\cdot10^{-5}$	&	0.7	\\\hline
Twin-1	&	$10^{10}$	&	0.13	&	0.133	&	$1.58\cdot10^{-16}$	&	$1\cdot10^{-16}$	&	$1\cdot10^{-10}$	\\
Twin-2	&	$10^4$	&	0.131	&	0.131	&	$1.59\cdot10^{-4}$	&	$1\cdot10^{-5}$	&	0.1	\\
\hline
\end{tabular}
\end{center}
\caption{Definitions of the benchmark points for the SB model in terms of the free parameters $M$ (in GeV) and $\lambda_i$. The parameter values $v=246$ GeV, $\beta_1=0.0166$, $\sin^2\theta_W=0.23126$, $g_L=0.65170$, $\theta_W^\prime=0.62$, $g_R = g_L\tan\theta_W/\sin\theta_W^\prime$, $2g' = g_L\tan\theta_W/\cos\theta_W^\prime$ are kept fixed.}\label{t:benchmarkpoints}
\end{table}

\begin{table}
\begin{center}
\begin{tabular}{|l|c|c|c|c|}
\hline
Benchmark point	&	$m_{H_1^0}$	&	$m_{A^0}$	&	$m_{H^\pm}$	&	$m_{H_2^0}$	\\\hline\hline
SLH-1	&	$3.2\cdot10^9$	&	$3.2\cdot10^9$	&	$3.2\cdot10^9$	&	$6.9\cdot10^9$	\\
SLH-2	&	$3.2\cdot10^3$	&	$3.2\cdot10^3$	&	$3.2\cdot10^3$	&	$6.9\cdot10^3$	\\\hline
2HDM-1	&	488	&	488	&	500	&	$6.6\cdot10^9$	\\
2HDM-2	&	488	&	488	&	500	&	$6.6\cdot10^3$	\\\hline
LF-1	&	$3.9\cdot10^6$	&	$3.9\cdot10^6$	&	$3.9\cdot10^6$	&	447	\\
LF-2	&	$5.5\cdot10^3$	&	$5.5\cdot10^3$	&	$5.5\cdot10^3$	&	448	\\\hline
Compressed-1	&	496	&	496	&	500	&	332	\\
Compressed-2	&	496	&	496	&	500	&	334	\\\hline
VLF-1	&	$1.0\cdot10^6$	&	$1.0\cdot10^6$	&	$1.0\cdot10^6$	&	1.0	\\
VLF-2	&	$5.9\cdot10^3$	&	$5.9\cdot10^3$	&	$5.9\cdot10^3$	&	0.9	\\\hline
Twin-1	&	$7.1\cdot10^4$	&	$7.1\cdot10^4$	&	$7.1\cdot10^4$	&	126	\\
Twin-2	&	$2.2\cdot10^3$	&	$2.2\cdot10^3$	&	$2.2\cdot10^3$	&	126	\\
\hline
\end{tabular}
\end{center}
\caption{Scalar masses for the benchmark points defined in \cref{t:benchmarkpoints}. All masses are given in GeV. The mass of the SM-like Higgs $h^0$ has been tuned to 126 GeV in each case.}\label{t:benchmarkMasses}
\end{table}

\subsection{The LET model}\label{s:completeLETbenchmarks}
In the SB model, the new physics decouples from the SM unless some of the dimensionless scalar parameters are set to $\ord{\xi^2}$ values.
The reason is the fact that there is a large hierarchy $v\ll M$ among the vev parameters of $\Phi_1$, $\Phi_R$.
The picture changes with the inclusion of the bidoublet $\Phi_2$, since its vev parameters are bounded by the electroweak scale.
In the absence of a large hierarchy between the vev components of $\Phi_1$ and $\Phi_2$, we expect that the LET model allows for large mixing between the components of both bidoublets if the dimensionless scalar-potential parameters have $\ord{1}$ values.
Hence the model naturally contains new scalar particles with $\ord{v}$ masses.
As in the SB model, significant mixing of $\Phi_1$, $\Phi_2$ with the components of $\Phi_R$ is only expected for unnaturally small values of some of the parameters.

These insights have important consequences for the benchmark scenarios considered in the previous section.
The SLH scenarios have no analogon in the LET model: there is always new physics within experimental reach.
The 2HDM-like, LF, VLF and Twin scenarios become natural possibilities, since the 2HDM-like scalars can naturally have $\ord{v}$ masses.
Moreover, mixing of $\Phi_1$ and $\Phi_2$ may make these scalars fermiophobic, in which case this scenario could be distinguished from the usual 2HDM.
Hence we expect the LET model to be predictive: it allows for phenomenologically interesting, experimentally testable scenarios, since the new physics does not decouple from the SM in the large-$M$ limit.

In the following, we discuss the prospects for measuring the Higgs-coupling modifications in the context of the LET model.
To this end, we define a new set of benchmark points, inspired by the considerations given above.
As a starting point for choosing the parameter values, we observe the following about the scalar masses in~\cref{eq:completeLETscalarMasses}.
The main contribution to $m_{h^0}$ is given by $\lambda_1+\lambda_2c^2_{\beta_1}$, with an overall scaling factor $c^2_\alpha$ due to the presence of the second bidoublet.
This means that a smaller $\alpha$ is generally accompanied by a smaller value for $\lambda_1+\lambda_2c^2_{\beta_1}$.
Similarly, $m_{H^0_1}$ is mainly determined by $\widetilde\lambda_1+\widetilde\lambda_2c^2_{\beta_2}$ with an overall factor $s^2_{\alpha}$, so smaller values of $\alpha$ should be compensated by larger values of $\widetilde\lambda_1+\widetilde\lambda_2c^2_{\beta_2}$.
The $h^0-H^0_1$ mass difference and mixing are governed by $s_{(2\alpha)}$ as well as the scalar parameters $\widetilde\lambda_{5,6,7,8,9,10}$, so we tune these parameters until we have a parameter set that corresponds to the desired benchmark scenario.
The scalar $H^0_4$ has $m_{H^0_4} \approx \sqrt{\lambda_3}M$, so we should take $\lambda_3 > 0$.
The squared masses of $A^0_1$, $H^\pm_1$ are determined by $\widetilde\lambda_{6,7,9,10}$ with an overall minus sign, so we take these parameters to be negative to guarantee a positive-definite mass matrix.
Positivity of the squared masses of $H^0_2$, $A^0_2$, $H^\pm_2$ requires $\lambda_5 > 0$, and the squared masses of $H^0_3$, $A^0_3$, $H^\pm_3$ require $\widetilde\lambda_4/c_{(2\beta_2)} > 0$.
We tune the parameters such that $m_{h^0} = 126$ GeV is fixed and enforce a set of vacuum stability conditions.
The parameter values for the benchmark points are given in \cref{t:completeLETbenchmarkPoints}, and the resulting scalar masses are given in \cref{t:completeLETscalarMasses}.

\begin{table}
{\renewcommand{\arraystretch}{1.3}
\begin{center}
\begin{tabular}{|c|c|c|c|c|c|}
\hline
	&	2HDM-3	&	2HDM-4	&	VLF-3	&	Twin-3	&	Twin-4	\\\hline\hline
$\sin\alpha$	&	0.93	&	0.43	&	0.50	&	0.72	&	0.33	\\
$\sin\beta_2$	&	0.17	&	0.65	&	0.17	&	0.16	&	0.11	\\\hline
$\lambda_1$	&	1.0	&	0.20	&	0.13	&	0.34	&	0.17	\\
$\lambda_2$	&	1.0	&	0.18	&	0.13	&	0.35	&	0.16	\\
$\lambda_3$	&	0.50	&	0.50	&	0.50	&	0.50	&	0.42	\\
$\lambda_4$	&	0.010	&	0.12	&	0.13	&	0.27	&	0.12	\\
$\lambda_5$	&	0.20	&	0.50	&	0.20	&	0.20	&	0.50	\\\hline
$\widetilde\lambda_1$	&	0.40	&	1.3	&	0.27	&	0.34	&	1.3	\\
$\widetilde\lambda_2$	&	0.40	&	1.3	&	0.27	&	0.34	&	1.2	\\
$\widetilde\lambda_3$	&	0.27	&	0.20	&	0.27	&	0.27	&	0.19	\\
$\widetilde\lambda_4$	&	0.20	&	0.10	&	0.20	&	0.20	&	1.0	\\
$\widetilde\lambda_5$	&	0.20	&	0.046	&	0.54	&	0.32	&	0.060	\\
$\widetilde\lambda_6$	&	-0.40	&	-0.50	&	-0.43	&	-0.42	&	-0.30	\\
$\widetilde\lambda_7$	&	-0.24	&	-0.20	&	-0.24	&	-0.24	&	-0.30	\\
$\widetilde\lambda_8$	&	0.84	&	0.95	&	0.83	&	0.84	&	1.0	\\
$\widetilde\lambda_9$	&	-0.050	&	-0.10	&	-0.04	&	-0.053	&	-0.10	\\
$\widetilde\lambda_{10}$	&	-0.30	&	-0.30	&	-0.30	&	-0.30	&	-0.30	\\\hline
\end{tabular}
\end{center}}
\caption{Benchmark-point definitions for the LET model. The values $v = 246$ GeV, $M = 10^{10}$ GeV, $\beta_1 = 0.0166$ are kept fixed.}\label{t:completeLETbenchmarkPoints}
\end{table}

\begin{table}
{\renewcommand{\arraystretch}{1.3}
\begin{center}
\begin{tabular}{|c|c|c|c|c|c|}
\hline
	&	2HDM-3	&	2HDM-4	&	VLF-3	&	Twin-3	&	Twin-4	\\\hline\hline
$m_{h^0}$	&	126	&	126	&	126	&	126	&	126	\\
$m_{H^0_1}$	&	182	&	148	&	3.9	&	126	&	126	\\
$m_{H^0_2}$	&	$3.2\cdot10^9$	&	$5.0\cdot10^9$	&	$3.2\cdot10^9$	&	$3.2\cdot10^9$	&	$5.0\cdot10^9$	\\
$m_{H^0_3}$	&	$3.3\cdot10^9$	&	$5.8\cdot10^9$	&	$3.3\cdot10^9$	&	$3.2\cdot10^9$	&	$7.2\cdot10^9$	\\
$m_{H^0_4}$	&	$7.1\cdot10^9$	&	$7.1\cdot10^9$	&	$7.1\cdot10^9$	&	$7.1\cdot10^9$	&	$6.5\cdot10^9$	\\\hline
$m_{A^0_1}$	&	179	&	134	&	179	&	179	&	190	\\
$m_{A^0_2}$	&	$3.2\cdot10^9$	&	$5.0\cdot10^9$	&	$3.2\cdot10^9$	&	$3.2\cdot10^9$	&	$5.0\cdot10^9$	\\
$m_{A^0_3}$	&	$3.3\cdot10^9$	&	$5.8\cdot10^9$	&	$3.3\cdot10^9$	&	$3.2\cdot10^9$	&	$7.2\cdot10^9$	\\\hline
$m_{H^\pm_1}$	&	171	&	135	&	173	&	173	&	173	\\
$m_{H^\pm_2}$	&	$3.2\cdot10^9$	&	$5.0\cdot10^9$	&	$3.2\cdot10^9$	&	$3.2\cdot10^9$	&	$5.0\cdot10^9$	\\
$m_{H^\pm_3}$	&	$3.3\cdot10^9$	&	$5.8\cdot10^9$	&	$3.3\cdot10^9$	&	$3.2\cdot10^9$	&	$7.2\cdot10^9$	\\\hline
\end{tabular}
\end{center}}
\caption{Scalar masses for each of the benchmark points defined in \cref{t:completeLETbenchmarkPoints}. All masses are in GeV.}\label{t:completeLETscalarMasses}
\end{table}

We have defined two 2HDM-like benchmarks, in which the scalars $H_1^0$, $A_1^0$, $H^\pm_1$ all have masses in the $\ord{100\text{ GeV}}$ range.
In 2HDM-3, the state $h^0$ is almost purely $h^0_{1,11}$, whereas $H_1^0$ is 97\% $h^0_{2,11}$ and 3\% $h^0_{2,22}$.
The state $A_1^0$ ($H_1^\pm$) is 87\% $a^0_{1,11}$ ($h^\pm_{1,21}$) and 13\% $a^0_{2,11}$ ($h^\pm_{2,21}$).
The mixings are even larger in the 2HDM-4 benchmark: $h^0$ is 74\% $h^0_{1,11}$, 15\% $h^0_{2,11}$, and 11\% $h^0_{2,22}$ whereas $H^0_1$ is 25\% $h^0_{1,11}$, 43\% $h^0_{2,11}$ and 32\% $h^0_{2,22}$.
The state $A_1^0$ ($H_1^\pm$) is 18\% $a^0_{1,11}$ ($h^\pm_{1,21}$), 47\% $a^0_{2,11}$ ($h^\pm_{2,21}$), and 35\% $a^0_{2,22}$ ($h^\pm_{2,12}$).
In both cases, the 2HDM-like scalars have suppressed couplings to fermions with respect to the type-II 2HDM.

We also define the VLF-3 scenario, in which $H^0_1$ has a mass of only a few GeV.
The states $A^0_1$, $H^\pm_1$ have masses within experimental reach.
The scalar mixing is significant: $h^0$ is 65\% $h^0_{1,11}$, 34\% $h^0_{2,11}$, and 1\% $h^0_{2,22}$, so we expect reduced fermion couplings.
The state $H^0_1$ is 35\% $h^0_{1,11}$, 63\% $h^0_{2,11}$, and 2\% $h^0_{2,22}$.
Since it is light and mainly fermiophobic, it could have evaded the LEP searches.
The state $A_1^0$ ($H_1^\pm$) is 25\% $a^0_{1,11}$ ($h^\pm_{1,21}$), 73\% $a^0_{2,11}$ ($h^\pm_{2,21}$), and 2\% $a^0_{2,22}$ ($h^\pm_{2,12}$).

Furthermore, we define two Twin benchmarks with different amounts of scalar mixing.
For Twin-3, the state $h^0$ is 87\% $h^0_{1,11}$ and 13\% $h^0_{2,11}$, whereas $H^0_1$ is 13\% $h^0_{1,11}$, 85\% $h^0_{2,11}$, and 2\% $h^0_{2,22}$.
The lightest $CP$-odd and charged states are almost 50-50 mixtures of fermiophilic and fermiophobic states: $A^0_1$ ($H^\pm_1$) is 51\% $a^0_{1,11}$ ($h^\pm_{1,21}$), 47\% $a^0_{2,11}$ ($h^\pm_{2,21}$), and 1\% $a^0_{2,22}$ ($h^\pm_{2,12}$).
For Twin-4, $h^0$ is almost purely $h^0_{1,11}$ whereas $H^0_1$ is 99\% $h^0_{2,11}$ and 1\% $h^0_{2,22}$.
The lightest $CP$-odd and charged states are mostly fermiophobic: $A^0_1$ ($H^\pm_1$) is 11\% $a^0_{1,11}$ ($h^\pm_{1,21}$), 88\% $a^0_{2,11}$ ($h^\pm_{2,21}$), and 1\% $a^0_{2,22}$ ($h^\pm_{2,12}$).

\section{Higgs-coupling modifications}\label{s:HCMs}
\begin{figure}[t]
\begin{center}
\includegraphics[width=.43\textwidth]{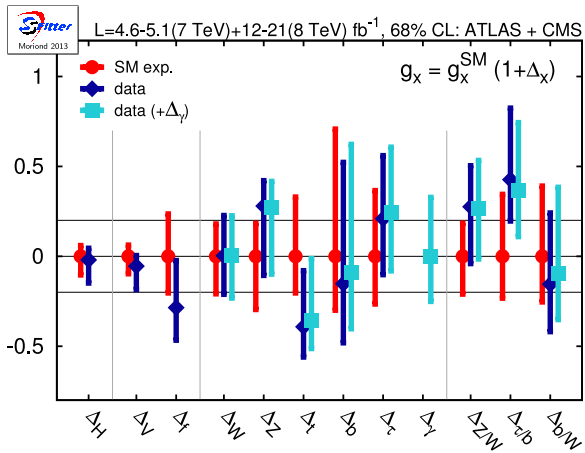}
\end{center}
\caption{Most recent fit of the Higgs-coupling modifications to LHC data. The red points correspond to the expected SM result $\Delta_x = 0$, whereas the dark blue points give the results from the data if the photon coupling is assumed to be determined by the $W$ and $t$ loops only. The light blue points give the results if a free coupling shift in $\Delta_\gamma$ due to new physics is allowed. Figure taken from ref.~\cite{Lopez-Val:2013yba}.}\label{f:Higgscouplingmeasurements}
\end{figure}

In the SM, the Higgs couplings are fixed in terms of the particle masses and the vev of the Higgs field.
Hence an independent measurement of these couplings provides an important test of the SM.
These couplings are generally modified in the presence of an extended Higgs sector \cite{Lopez-Val:2013yba}.
The Higgs-coupling modifications $\Delta_x$ are defined as the deviations of the Higgs couplings $g_x\equiv g_{h^0xx}$ from their SM values, where $x$ is any SM particle and $g_{h^0xx}$ is the coefficient of the operator $h^0xx$ in the Lagrangian:
\begin{equation}
g_x = (1 + \Delta_x)g_x^\text{SM}. \label{eq:CMdefinition}
\end{equation}
The loop-induced Higgs coupling to photons can be written as follows:
\begin{equation}
g_\gamma = (1 + \Delta_\gamma^\text{SM} + \Delta_\gamma)g_\gamma^\text{SM}.
\end{equation}
Here $\Delta_\gamma^\text{SM}$ is the coupling modification that is induced by coupling modifications of the Standard-Model particles generating the coupling.
The term $\Delta_\gamma$ represents contributions from non-SM particles running in the loops.
The Higgs-coupling modifications have been extracted from LHC data using the tool SFitter \cite{Lafaye:2009vr,Klute:2012pu,Plehn:2012iz,Lopez-Val:2013yba} (see \cref{f:Higgscouplingmeasurements}).

\subsection{The SB model}
The Higgs-coupling modifications of the SB model are given in \cref{a:simplifiedLEThcms}.
In order to see whether our benchmark scenarios could be distinguished from the SM in experiment, we calculate the Higgs-coupling modifications numerically for each benchmark point.
The results are listed in \cref{t:benchmarkCMs}.

\begin{table*}
\begin{tabular}{|l|c|c|c|c|c|c|c|}
\hline
Benchmark point	&	$\Delta_W$	&	$\Delta_Z$	&	$\Delta_t$	&	$\Delta_b$	&	$\Delta_\gamma$	&	$\Delta_{\lambda_{3h}}$	&	$\Delta_{\lambda_{4h}}$	\\\hline\hline
SLH-1	&	$-1.4\cdot10^{-16}$	&	0.0	&	$-1.4\cdot10^{-16}$	&	0.0	&	$-5.0\cdot10^{-17}$	&	0.0	&	0.83	\\
SLH-2	&	$-1.4\cdot10^{-4}$	&	$-4.1\cdot10^{-4}$	&	$-1.4\cdot10^{-4}$	&	$2.1\cdot10^{-3}$	&	$-5.0\cdot10^{-5}$	&	$-4.2\cdot10^{-4}$	&	0.83	\\\hline
2HDM-1	&	$-1.6\cdot10^{-6}$	&	$-1.6\cdot10^{-6}$	&	$-3.2\cdot10^{-5}$	&	0.11	&	$-1.7\cdot10^{-3}$	&	$-9.7\cdot10^{-5}$	&	2.1	\\
2HDM-2	&	$-3.8\cdot10^{-4}$	&	$-6.4\cdot10^{-4}$	&	$-4.1\cdot10^{-4}$	&	0.11	&	$-1.6\cdot10^{-3}$	&	$-1.2\cdot10^{-3}$	&	2.1	\\\hline
LF-1	&	$-6.2\cdot10^{-7}$	&	$-6.2\cdot10^{-7}$	&	$-6.2\cdot10^{-7}$	&	$-6.2\cdot10^{-7}$	&	$-2.6\cdot10^{-11}$	&	$-1.9\cdot10^{-6}$	&	$1.2\cdot10^{-5}$	\\
LF-2	&	$-2.9\cdot10^{-3}$	&	$-3.1\cdot10^{-3}$	&	$-2.9\cdot10^{-3}$	&	$1.1\cdot10^{-3}$	&	$5.1\cdot10^{-5}$	&	$-8.6\cdot10^{-3}$	&	$5.5\cdot10^{-2}$	\\\hline
Compressed-1	&	$-1.6\cdot10^{-7}$	&	$-1.6\cdot10^{-7}$	&	$-9.6\cdot10^{-6}$	&	$3.4\cdot10^{-2}$	&	$-1.6\cdot10^{-3}$	&	$-9.6\cdot10^{-6}$	&	$-1.4\cdot10^{-5}$	\\
Compressed-2	&	$-9.1\cdot10^{-3}$	&	$-9.4\cdot10^{-3}$	&	$-9.1\cdot10^{-3}$	&	$3.5\cdot10^{-2}$	&	$-1.5\cdot10^{-3}$	&	$-2.7\cdot10^{-2}$	&	$7.1\cdot10^{-2}$	\\\hline
VLF-1	&	$-3.6\cdot10^{-7}$	&	$-3.6\cdot10^{-7}$	&	$-3.6\cdot10^{-7}$	&	$-3.6\cdot10^{-7}$	&	$-4.1\cdot10^{-10}$	&	$-1.1\cdot10^{-6}$	&	$-2.2\cdot10^{-6}$	\\
VLF-2	&	$-6.6\cdot10^{-4}$	&	$-9.3\cdot10^{-4}$	&	$-6.6\cdot10^{-4}$	&	$-2.2\cdot10^{-3}$	&	$-8.8\cdot10^{-5}$	&	$-2.0\cdot10^{-3}$	&	$-4.0\cdot10^{-3}$	\\\hline
Twin-1	&	$-1.9\cdot10^{-7}$	&	$-1.9\cdot10^{-7}$	&	$-1.9\cdot10^{-7}$	&	$1.4\cdot10^{-6}$	&	$-8.1\cdot10^{-8}$	&	$-5.6\cdot10^{-7}$	&	$-7.5\cdot10^{-7}$	\\
Twin-2	&	-0.21	&	-0.21	&	-0.21	&	-0.18	&	$6.9\cdot10^{-4}$	&	$-0.52$	&	$-0.61$	\\
\hline
\end{tabular}
\caption{Numerical results for the Higgs-coupling modifications for the benchmark points defined in \cref{t:benchmarkpoints}.}\label{t:benchmarkCMs}
\end{table*}

For the SLH benchmarks, the modification of the quartic Higgs self-coupling is as large as 83\%, whereas the other couplings have negligible deviations from their SM values.
Hence if we could measure both Higgs self-couplings, a SB model with a large hierarchy could be distinguished from the SM by the strength of the quartic Higgs self-coupling, even if $M$ is large.
Like in the 2HDM \cite{Osland:2008aw}, the self-couplings are generally modified due to the more complicated structure of the scalar potential, although these modifications may also vanish along some directions of the parameter space (see \cref{a:simplifiedLEThcms}).

For the 2HDM-like benchmarks, there is an 11\% increase of the Higgs coupling to $b$ quarks.
The reason is that the main contribution to $\Delta_b$ in this scenario is proportional to $\xi^2/\lambda_5 \sim 0.1$.
The $W,Z$ and $t$ couplings obtain similar modifications, but these are suppressed by factors $s^2_{(4\beta_1)} \sim 10^{-3}$ and $s^2_{\beta_1} \sim 10^{-4}$ respectively.
Hence the 2HDM-like hierarchy is characterized by an increase in only the $b$ coupling.
The quartic Higgs self-coupling is enhanced by a factor 3.

The LF, Compressed, and VLF benchmarks have coupling modifications at or below the percent level.
These scenarios would not be distinguishable from the SM via the Higgs-coupling modifications.
The Twin-1 benchmark has very small coupling modifications as well, but the Twin-2 benchmark shows 20\% reductions of all tree-level Higgs couplings and 50-60\% decreases of both self-couplings.
Currently, the errors on the Higgs couplings are still large enough to allow a 20\% deviation (see \cref{f:Higgscouplingmeasurements}).
If the errors can be reduced after the 14 TeV run, the Twin-2 benchmark can be put to the test.

\subsection{The LET model}
Because of the large number of free parameters and scalar mixing angles, we restrict ourselves to a numerical analysis of the Higgs-coupling modifications for the benchmark points defined in \cref{s:completeLETbenchmarks}.
The resulting coupling modifications have been summarized in \cref{t:completeLEThiggsCouplingModifications}.
Note that contrary to the SB model, the photon-coupling modification is not negligibly small any more: the charged scalar $H^\pm_1$ has a mass of order $v$ and hence yields a sizable contribution to the effective photon coupling.

\begin{table}[t]
{\renewcommand{\arraystretch}{1.3}
\begin{center}
\begin{tabular}{|c|c|c|c|c|c|}
\hline
	&	2HDM-3	&	2HDM-4	&	VLF-3	&	Twin-3	&	Twin-4	\\\hline\hline
$\Delta_W$	&	-0.69	&	-0.01	&	-0.01	&	-0.61	&	-0.06	\\
$\Delta_Z$	&	-0.69	&	-0.001	&	-0.001	&	-0.60	&	-0.05	\\
$\Delta_t$	&	1.8	&	-0.05	&	-0.07	&	0.34	&	0.06	\\
$\Delta_b$	&	1.8	&	-0.05	&	-0.07	&	0.34	&	0.06	\\
$\Delta_\gamma$	&	-0.05	&	-0.09	&	-0.07	&	-0.03	&	-0.05	\\\hline
\end{tabular}
\end{center}}
\caption{Higgs-coupling modifications for the benchmark points of the LET model, defined in \cref{t:completeLETbenchmarkPoints}.}\label{t:completeLEThiggsCouplingModifications}
\end{table}

The 2HDM-3 benchmark has large coupling modifications: the couplings to $V=W, Z$ are suppressed by a factor 0.3, whereas the couplings to $t$, $b$ are enhanced by almost a factor 3.
This is not surprising: the $W$, $Z$ couplings of the SB model are proportional to $v$, whereas the $t$, $b$ couplings are proportional to $m_{t,b}/v$.
In the LET model, we have to substitute $v\rightarrow v\cos\alpha$ (see \cref{eq:completeLETvevRedefinition}).
Since $h^0$ is almost purely $h^0_{1,11}$ in the 2HDM-3 benchmark, the $W$, $Z$ couplings are suppressed by $\cos\alpha = 0.36$ whereas the $t$, $b$ couplings are enhanced by $1/\cos\alpha = 2.8$.
This benchmark point is clearly incompatible with the measured coupling modifications in \cref{f:Higgscouplingmeasurements}.
In contrast, the 2HDM-4 benchmark has smaller but still sizable coupling modifications.
The $V$ couplings are suppressed by $\cos\alpha = 0.90$, but the total coupling is a few percent higher because of the contribution from the second bidoublet.
On the other hand, the quark couplings are enhanced by a factor $1/\cos\alpha = 1.11$, but the total coupling modifications are negative: since $h^0$ contains a significant admixture of the fermiophobic $\Phi_2$, the $t$, $b$ couplings are reduced.
The coupling modifications for 2HDM-4 are consistent with the measured coupling modifications.

The VLF-3 scenario has large mixing between the fermiophilic and fermiophobic scalar gauge eigenstates.
As such, there is again a tension between $\cos\alpha$ and $\Phi_1-\Phi_2$ mixing.
The resulting coupling modifications are at the percent level, all compatible with the measured values.

Like the 2HDM-3 benchmark, the Twin-3 scenario has large coupling modifications.
The $W$, $Z$ couplings are reduced by about 60\%, mostly due to interference between the contributions of the fermiophobic and fermiophilic scalar components.
The $t$, $b$ couplings are enhanced by 34\%: the factor $1/\cos\alpha = 1.4$ is slightly reduced by $\Phi_1-\Phi_2$ mixing.
This benchmark point is incompatible with the data.
However, the Twin-4 scenario has percent-level coupling modifications, compatible with the measured values.
This is because the $V$ couplings are suppressed by $\cos\alpha = 0.94$ and the fermion couplings are enhanced by $1/\cos\alpha = 1.06$, and $\Phi_1-\Phi_2$ mixing is negligible.

We have illustrated that the LET model is predictive and allows for various interesting phenomenological scenarios.
The model allows for large coupling modifications as well as moderate ones that can be expected to be measurable, hence the model is testable.
A more thorough analysis of the parameter space is required to see which parameter values are preferred by experiment.

\section{Very light fermiophobic scalars}\label{s:VLF}
The VLF scenario contains a fermiophobic scalar particle with a mass of $\ord{1\text{ GeV}}$ in addition to the SM-like Higgs.
Such a particle would only decay into pairs of photons, and is not necessarily ruled out since it could have escaped detection so far.
We now review the experimental bounds that are relevant to this scenario.

Fermiophobic Higgs particles are not unique to the SB model: they also appear in a type-I 2HDM with $\alpha = \pi/2$ \cite{Akeroyd:1995hg} and in models with $SU(2)_L$-triplet Higgs fields \cite{Gunion:1989ci,Bamert:1993ah}.
Mass bounds from direct searches are readily available in the literature \cite{Agashe:2014kda}.
Assuming SM cross sections, the four LEP experiments \cite{Abreu:2001ib,Heister:2002ub,Achard:2002jh,Abbiendi:2002yc} have placed a lower limit $m_H > 107\text{ GeV}$ on the mass of a fermiophobic Higgs by looking for decays into pairs of photons.
More recent searches by ATLAS \cite{Aad:2012yq} and CMS \cite{Chatrchyan:2013sfs} in the diphoton channel as well as the $WW$, $ZZ$ channels \cite{Chatrchyan:2012vva} extend this lower limit to $m_H > 194 \text{ GeV}$.
However, the cuts on the energy of the photon pair in the LEP analyses make these searches insensitive to fermiophobic Higgs particles with masses below 10 GeV.

In order to see to what extent the lower mass bound applies to the fermiophobic Higgs $H_2^0$ of the SB model, we need to examine its couplings to SM particles.
The field $\Phi_R$ has no tree-level couplings to fermions, and its neutral component is a SM singlet.
Hence $H_2^0$ only couples to the SM through $W-W'$ mixing, $Z-Z'$ mixing and scalar mixing.
The former two are negligible since they are of order $\xi^2$.
Scalar mixing can become substantial in the Twin scenario, but it is negligible in the VLF scenario.
Hence $H_2^0$ has only $\ord{\xi^2}$ couplings to the SM in the SB model, and therefore the experimental bounds are evaded trivially.

This may change in the LET model.
We have seen that a very light fermiophobic Higgs with a mass of a few GeV becomes a natural possibility.
Since it is $\Phi_2$-like, it has significant couplings to $W$, $Z$, distinguishing it from the very light fermiophobic Higgs of the SB model, distinguishing it from the very light fermiophobic Higgs of the SB model.
Hence such a state could have significant production rates at the LHC.
A more thorough analysis of the production cross-section is necessary in order to predict the signal strength, which is beyond the scope of this work.

\section{A Twin Higgs scenario}\label{s:twinHiggs}
The Higgs signal strength in each channel has been measured at the LHC \cite{Aad:2013wqa,Chatrchyan:2012ufa,Chatrchyan:2014nva}.
These measurements constrain the Twin scenario.
In order to compare the Twin benchmarks to these experimental results, we have to consider the production cross-sections and branching ratios for both $h^0$ and $H_2^0$.

We parameterize the twin states $S^0 = h^0, H^0_2$ in terms of the gauge eigenstates as
\begin{align}
h^0 =& a_1h^0_{1,11} + a_2h^0_{1,22} + a_3h^0_R, \notag\\
H_2^0 =& a_1^\text{FP}h^0_{1,11} + a_2^\text{FP}h^0_{1,22} + a_3^\text{FP}h^0_R, \label{eq:twinMixingCoefficients}
\end{align}
where FP stands for fermiophobic.
The coefficients for the Twin-1 benchmark are given by
\begin{align}
(a_1, a_2, a_3) =& (0.9999, 0.0166, 0.0006),\quad (\text{Twin-1}) \notag\\
(a_1^\text{FP}, a_2^\text{FP}, a_3^\text{FP}) =& (-0.0006, -0.00001, 0.9999), \label{eq:Twin1coefficients}
\end{align}
whereas for the Twin-2 benchmark they are
\begin{align}
(a_1, a_2, a_3) =& (0.7874, 0.0136, -0.6162),\quad (\text{Twin-2}) \notag\\
(a_1^\text{FP}, a_2^\text{FP}, a_3^\text{FP}) =& (0.6162, 0.0096, 0.7876). \label{eq:Twin2coefficients}
\end{align}
We need to find the signal strengths $\mu_x(S^0)$ for each decay channel $S^0 \rightarrow xx$:
\begin{equation}
\mu_x(S^0) = \frac{\sigma(pp\rightarrow S^0)\times BR(S^0 \rightarrow xx)}{\sigma(pp\rightarrow h^0)_\text{SM} \times BR(h^0 \rightarrow xx)_\text{SM}},
\end{equation}
where $\sigma(pp\rightarrow S^0)$ is the production cross-section for $S^0$.
Then we need to add the signal strengths of $h^0$ and $H^0_2$.
In order to estimate the magnitude of the deviations from their SM values, we neglect loop corrections in the following discussion.

At the LHC, the Higgs can be produced in vector-boson fusion, VH associated production, gluon fusion, and production in association with $t\bar{t}$ pairs \cite{Dittmaier:2011ti}.
The former two processes are proportional to the Higgs coupling to vector bosons, whereas the latter two scale with the top coupling (the main contribution to gluon fusion comes from a top-quark loop).
For $h^0$, we have $\Delta_1 \equiv \Delta_W = \Delta_Z = \Delta_t = a_1 - 1$ (see \cref{a:simplifiedLEThcms}, we neglect $\ord{\xi^2}$ and $\ord{\beta_1}$ corrections).
Hence as a tree-level approximation we have
\begin{equation}
\frac{\sigma(pp \rightarrow h^0)}{\sigma(pp \rightarrow h^0)_\text{SM}} = a_1^2. \label{eq:h0crosssection}
\end{equation}
Similarly, $\Delta_1^\text{FP} \equiv \Delta_W^\text{FP} = \Delta_Z^\text{FP} = \Delta_t^\text{FP} = a_1^\text{FP} - 1$, up to $\ord{\xi^2}$ and $\ord{\beta_1}$ corrections.
Hence at tree level we have
\begin{equation}
\frac{\sigma(pp \rightarrow H_2^0)}{\sigma(pp \rightarrow h^0)_\text{SM}} = (a_1^\text{FP})^2. \label{eq:H20crosssection}
\end{equation}
For the Twin-1 benchmark we have $a_1^2 = 1$, $(a_1^\text{FP})^2 = 4\cdot10^{-7}$ (see \cref{eq:Twin1coefficients}).
That is, $h^0$ is produced at the same rate as the SM Higgs, whereas $H^0_2$ production is suppressed.
In the Twin-2 benchmark, however, both states have a significant production rate since $a_1^2 = 0.62$, $(a_1^\text{FP})^2 = 0.38$.

As for the branching ratios, we only take into account the decay channels listed in \cref{t:branchingratios}; all other channels have negligibly small branching ratios.
The given SM values were calculated with HDecay \cite{Djouadi:1997yw} using $m_h = 126$ GeV.
We estimate the corresponding branching ratios of the SB model using the Higgs-coupling modifications. According to \cref{eq:yukawalag} the $b$ quark and the $\tau$ couple to the same scalar gauge eigenstate, namely $h^0_{1,22}$.
We thus assume that $\Delta_\tau = \Delta_b \equiv \Delta_2$ and $\Delta_\tau^\text{FP} = \Delta_b^\text{FP} \equiv \Delta_2^\text{FP}$. 

Now we are ready to calculate the branching ratios of the SB model.
The partial decay widths for the $h^0$ and $H_2^0$ decays into $y_1y_1 = WW$, $ZZ$, $gg$, $\gamma\gamma$, $cc$ scale with respectively $(1+\Delta_1)^2$ and $(1+\Delta_1^\text{FP})^2$, whereas they scale with respectively $(1+\Delta_2)^2$ and $(1+\Delta_2^\text{FP})^2$ for the $y_2y_2 = bb$, $\tau\tau$ decay channels.
Thus the branching ratios for the SB model are given by
\begin{align}
BR(h^0 \rightarrow xx) =& \frac{(1+\Delta_x)^2 BR(h^0 \rightarrow xx)_\text{SM}}{(1+\Delta_1)^2 \widetilde{BR}_1 + (1+\Delta_2)^2 \widetilde{BR}_2}, \notag\\
BR(H^0_2 \rightarrow xx) =& \frac{(1+\Delta_x^\text{FP})^2 BR(h^0 \rightarrow xx)_\text{SM}}{(1+\Delta_1^\text{FP})^2 \widetilde{BR}_1 + (1+\Delta_2^\text{FP})^2 \widetilde{BR}_2}, \notag\\
\widetilde{BR}_1 \equiv& \sum_{y_1 = W,Z,g,\gamma,c} BR(h^0 \rightarrow y_1y_1)_\text{SM}, \notag\\
\widetilde{BR}_2 \equiv& \sum_{y_2 = b,\tau} BR(h^0 \rightarrow y_2y_2)_\text{SM}. \label{eq:h0branchingratio}
\end{align}
In the Twin-1 benchmark we have $|\Delta_{1,2}| \ll 1$ (see \cref{t:benchmarkCMs}), hence the branching ratios for $h^0$ barely deviate from their SM values.
The branching ratios for $H^0_2$ are small, since $|1+\Delta_{1,2}^\text{FP}| = |a_{1,2}^\text{FP}| \ll 1$.
The branching ratios for the Twin-2 benchmark are listed in \cref{t:branchingratios}.

The total signal strength is given by the sum of the contributions from $h^0$ and $H_2^0$.
Combining \cref{eq:h0crosssection,eq:H20crosssection,eq:h0branchingratio}, we find
\begin{align}
\mu_{x,\text{tot}} =& \frac{a_1^2 (1 + \Delta_x)^2}{(1 + \Delta_1)^2 \widetilde{BR}_1 + (1 + \Delta_2)^2 \widetilde{BR}_2} \notag\\
&+ \frac{(a_1^\text{FP})^2 (1 + \Delta_x^\text{FP})^2}{(1 + \Delta_1^\text{FP})^2 \widetilde{BR}_1 + (1 + \Delta_2^\text{FP})^2 \widetilde{BR}_2}.
\end{align}
We find $\mu_{x,tot} = 1$ for the Twin-1 benchmark: $h^0$ contributes with the SM strength whereas $H^0_2$, despite its identical mass, is hardly produced and cannot show itself by decays to SM final states.
The signal strengths for the Twin-2 benchmark have been summarized in \cref{t:branchingratios}. In this case two states with the same mass but different decay properties are present. Still,  none of the decay channels has a total signal strength that deviates significantly from 1. 
The reason is  that the SM contributions to the signal strength are simply divided among the two scalars.


\begin{table}[t]
\begin{center}
\begin{tabular}{|c|c|c|c|c|c|c|c|}
\hline
&	$WW$	&	$ZZ$	&	$gg$	&	$\gamma\gamma$	&	$cc$	&	$bb$	&	$\tau\tau$	\\\hline\hline
$BR(h^0 \rightarrow xx)_\text{SM}$	&	0.216	&	0.027	&	0.077	&	0.002	&	0.026	&	0.594	&	0.057	\\\hline\hline
$BR(h^0 \rightarrow xx)$	&	0.206	&	0.026	&	0.073	&	0.002	&	0.025	&	0.610	&	0.059	\\
$BR(H_2^0\rightarrow xx)$	&	0.235	&	0.029	&	0.084	&	0.002	&	0.028	&	0.566	&	0.054	\\\hline\hline
$\mu_x(h^0)$	&	0.59	&	0.59	&	0.59	&	0.59	&	0.59	&	0.64	&	0.64	\\
$\mu_x(H_2^0)$	&	0.41	&	0.41	&	0.41	&	0.41	&	0.41	&	0.36	&	0.36	\\\hline
$\mu_x(h^0)+\mu_x(H^0_2)$	&	1.0	&	1.0	&	1.0	&	1.0	&	1.0	&	1.0	&	1.0	\\\hline
\end{tabular}
\end{center}
\caption{Branching ratios and signal strengths of $h^0$ and $H_2^0$ decays in the Twin-2 benchmark of the SB model. The SM values of the branching ratios were calculated with HDecay \cite{Djouadi:1997yw} for $m_h = 126$ GeV. We neglect the branching ratios for the $\mu\mu$, $ss$, $tt$, $Z\gamma$ decay channels.}\label{t:branchingratios}
\end{table}

The situation may change in the LET model, in which the fermiophobic scalar $H_1^0$ can be the twin partner of $h^0$.
The state $H_1^0$ is a mixture of components of both $\Phi_1$ and $\Phi_2$.
Since the latter is an $SU(2)_L$-antidoublet, it couples to $W$, $Z$.
After scalar mixing, these additional contributions can give the twin Higgs a total signal strength that differs from the SM prediction.
Note that the Higgs-coupling modifications of the LET model are not universal (see \cref{t:completeLEThiggsCouplingModifications}).
This means that the Higgs production cross-sections do not scale trivially, as they did in the SB model.
Hence a prediction of the twin-Higgs signal strength requires a more detailed analysis of the production cross-section.

\section{Conclusions}\label{s:conclusions}
In this work we have studied the low-energy phenomenology of the trinification model as described in refs.~\cite{Achiman:1978vg,Stech:2003sb,Stech:2008wd,Stech:2010gf,Stech:2012zr,Stech:2014tla}.
It is based on the trinification group $\trini$.
In order to simplify our study, we have integrated out the fields that obtain masses of the order of the trinification scale.
This resulted in a left-right-symmetric model with two scalar bidoublets $\Phi_1$, $\Phi_2$ and one right-handed doublet $\Phi_R$.
While the bidoublets obtain vevs of the order of the weak scale, the right-handed doublet has a vev $M$ that describes the scale at which the left-right symmetry is broken.
It may be as low as a few TeV but could also be much higher.
Only $\Phi_1$ couples to fermions.
We call this effective model the low-energy trinification (LET) model.
As an intermediate step towards a better understanding of the LET model, we have studied the Single-Bidoublet (SB) model, a simplified model in which $\Phi_2$ has been set to zero.

Our Ansatz for the Yukawa sector was based on the Yukawa Lagrangian of the trinification model.
The free parameters were used to fix the masses of the top and bottom quarks, which are the most important for comparison to experimental searches.
We left out a discussion of the first and second fermion generations.
An improved version of the LET model containing these lighter fermions would require the introduction of new Higgs fields, mixings with heavy states, and the consideration of renormalization-group effects.

In order to showcase the possible phenomenological scenarios of our model, we have defined a set of benchmark points for both the SB model and the LET model.
In the SB model, all new scalars decouple from the SM unless we tune some of the dimensionless scalar-coupling constants to $\ord{v^2/M^2}$ values.
For such small coupling constants, interesting phenomenological scenarios are possible at low energies, like a fermiophobic scalar particle with an $\ord{1\text{ GeV}}$ mass in addition to a SM-like Higgs, higher-mass states with fermiophobic components, or a degenerate state (`twin Higgs') at 126 GeV.
On the other hand, the full LET model always has at least one other $CP$-even scalar, one $CP$-odd scalar, and a pair of charged scalars with masses in the $\ord{100\text{ GeV}}$ range.
They appear without tuning the coupling constants to very small values.
The aforementioned phenomenological scenarios therefore become natural possibilities.

To show to what extent these scenarios can be distinguished from the SM in experiment, we have calculated the Higgs-coupling modifications for the benchmarks.
For the SB model, they are negligibly small in most cases.
However, the benchmarks of the LET model lead to sizable effects on the Higgs couplings.
Parts of the parameter space can already be excluded using the known limits given in the literature.

Subsequently, we have studied the scenario with a very light fermiophobic (VLF) Higgs in more detail.
Such a particle decays only into pairs of photons and escapes the currently available bounds from direct searches.
For the VLF Higgs of the SB model, the signal strength is very small since it has only $\ord{v^2/M^2}$ couplings to the SM.
We argued that the VLF Higgs of the LET model may have significant production rates at the LHC.
We also studied the twin Higgs scenario (degenerate Higgs state) in more detail.
For the SB model, we expect no significant deviations from the SM prediction even though each of the two degenerate states has different decay properties.
On the other hand, the twin Higgs of the LET model may lead to significant deviations, because the bidoublet $\Phi_2$ can introduce direct couplings of the new state to SM particles.
A more detailed analysis of the production cross-section appears necessary for a detailed comparison with the measured signal strengths.

We have calculated the phenomenological scenarios of the LET model at several benchmark points and found interesting consequences for the properties of the Higgs bosons.
Because of the number of coupling constants in the potential a systematic investigation of all regions of the parameter space is still missing.
Nevertheless, the examples given show the large variety of possibilities still not excluded by experiment.
The LET model turned out to be an interesting extension of the SM.
It is predictive and most properties can be tested or constrained using forthcoming LHC data.

\begin{acknowledgments}
The authors like to thank Tilman Plehn for fruitful and stimulating discussions.
JH thanks him in addition for supervising his PhD thesis and acknowledges financial support by the German Research Foundation (Deutsche Forschungsgemeinschaft).
\end{acknowledgments}

\appendix

\section{Gauge sector of the LET model}\label{a:gaugebosons}
The gauge-boson sector consists of seven fields: $W_{L,R}^{1,2,3}$ for $SU(2)_{L,R}$ and $B$ for $U(1)_{B-L}$.
The fields $W_L^3$, $W_R^3$, and $B$ are neutral, whereas the remaining fields mix to form the charge eigenstates $W_{L,R}^\pm \equiv (W_{L,R}^1 \mp iW_{L,R}^2)/\sqrt{2}$.
These charged states are rotated by an angle $\zeta$ into two pairs of charged mass eigenstates $W^\pm$ and $W^{\prime\pm}$:
\begin{equation}
\begin{pmatrix} W^\pm \\ W^{\prime\pm} \end{pmatrix}
= \begin{pmatrix} \cos\zeta & \sin\zeta \\ -\sin\zeta & \cos\zeta \end{pmatrix}
\begin{pmatrix} W_L^\pm \\ W_R^\pm \end{pmatrix}.
\end{equation}
Here the $W^\pm$ correspond to the charged vector bosons of the SM.
The $W^{\prime\pm}$ bosons are new massive vector bosons.
The mixing angle is very small (we define $s_x \equiv \sin{x}$, $c_x \equiv \cos{x}$ for the sake of brevity):
\begin{equation}
\zeta = \frac{g_L}{g_R}\xi^2(c^2_\alpha s_{(2\beta_1)} + s^2_\alpha s_{(2\beta_2)}) + \ord{\xi^4}. \label{eq:zeta}
\end{equation}
The masses of the charged mass eigenstates are given by
\begin{align}
m_W =& \frac{g_Lv}{2}\Big( 1 - \frac{\xi^2}{2}(c^2_\alpha s_{(2\beta_1)} + s^2_\alpha s_{(2\beta_2)})^2  + \ord{\xi^4} \Big), \notag\\
m_{W'} =& \frac{g_RM}{2}\left( 1 + \xi^2 + \ord{\xi^4} \right). \label{eq:Wmasses}
\end{align}
The three neutral gauge fields mix into a massless photon $A$, a massive $Z$ as in the SM, and a new massive state $Z'$.
These states are obtained through a rotation over three mixing angles $\theta_W$, $\theta_W^\prime$, $\eta$:
\begin{align}
A =& s_{\theta_W}W_L^3 + c_{\theta_W}(s_{\theta_W^\prime}W_R^3 + c_{\theta_W^\prime}B), \notag\\
Z =& c_{\theta_W}c_\eta W_L^3 + (c_{\theta_W^\prime}s_\eta - s_{\theta_W}s_{\theta_W^\prime}c_\eta)W_R^3 \notag\\
&- (s_{\theta_W}c_{\theta_W^\prime}c_\eta + s_{\theta_W^\prime}s_\eta)B, \notag\\
Z' =& -c_{\theta_W}s_\eta W_L^3 + (c_{\theta_W^\prime}c_\eta + s_{\theta_W}s_{\theta_W^\prime}s_\eta)W_R^3 \notag\\
&+ (s_{\theta_W}c_{\theta_W^\prime}s_\eta - s_{\theta_W^\prime}c_\eta)B.
\end{align}
The angle $\theta_W$ is the Weinberg angle; $\theta_W^\prime$ is an analogon of $\theta_W$ for the breaking of the left-right symmetry; and $\eta$ is the $Z-Z'$ mixing angle.
These angles are given in terms of the gauge couplings by
\begin{align}
\sin\theta_W =& \frac{2g'g_R}{\sqrt{4g^{\prime2}(g_L^2+g_R^2)+g_L^2g_R^2}},\quad \sin\theta_W^\prime = \frac{2g'}{\sqrt{g_R^2+4g^{\prime2}}}, \notag\\
\tan\eta =& \frac{g_R^2\sqrt{4g^{\prime2}(g_L^2+g_R^2)+g_L^2g_R^2}}{(g_R^2+4g^{\prime2})^2} \xi^2 + \ord{\xi^4}. \label{eq:neutralAngles}
\end{align}
The masses of the states $Z$, $Z'$ are given by
\begin{align}
m_Z =& \frac{g_Lv}{2\cos\theta_W} \left( 1 - \frac{\xi^2\cos^4\theta_W^\prime}{2} + \ord{\xi^4} \right), \notag\\
m_{Z'} =& \frac{g_RM}{2\cos\theta_W^\prime} \left( 1 + \frac{\xi^2\cos^4\theta_W^\prime}{2} + \ord{\xi^4} \right). \label{eq:Zmasses}
\end{align}
The electromagnetic coupling constant is found to be
\begin{equation}
e \equiv \frac{2g'g_Lg_R}{\sqrt{4g^{\prime2}(g_L^2+g_R^2)+g_L^2g_R^2}} = g_L\sin\theta_W. \label{eq:EMcoupling}
\end{equation}
Constraints on the $W'$ and $Z'$ masses as well as their mixing angles are widely available in the literature and allow us to constrain the new parameters $g_R$, $g'$, and $M$.
Direct searches for $W'$ and $Z'$ have been performed in various decay channels \cite{Abe:1997fd,*Abbiendi:2003dh,*Abdallah:2005ph,*Schael:2006wu,*Aad:2012dm,*Aad:2012ej,*Aad:2012gm,*Aad:2012hf,*Aad:2012vs,*Chatrchyan:2012gqa,*Chatrchyan:2012kk,*Chatrchyan:2012oaa,*Chatrchyan:2012rva,*CMS:2012yf,*Aad:2013wxa,*Chatrchyan:2013lga}, but mass bounds are only given under the assumption that the new vector bosons have the same couplings to fermions as $W$ and $Z$ respectively.
Hence they do not apply to the LET model.
More general constraints come from fits to electroweak precision data \cite{Chay:1998hd,delAguila:2010mx} and high-precision measurements \cite{Barenboim:1996nd,Bueno:2011fq,TWIST:2011aa}.
The right-handed coupling $g_R$ is constrained by the bound $g_R/g_L = 0.94 \pm 0.09$ from ref.~\cite{Barenboim:1996nd}.
Combining this bound with \cref{eq:EMcoupling} and the experimental values $g_L = 0.65170 \pm 0.00008$, $e = 0.313402 \pm 0.000017$ gives a constraint on $g'$ as well:
\begin{equation}
g_R = 0.61 \pm 0.06,\qquad g' = 0.22 \pm 0.01. \label{eq:gRgpBounds}
\end{equation}
The strongest constraint on $M$ comes from the bound $-0.00040 < \eta < 0.0026$ from ref.~\cite{Chay:1998hd} on the $Z-Z'$ mixing angle.
Combining it with \cref{eq:neutralAngles,eq:gRgpBounds} we find
\begin{equation}
M > 3.6\text{ TeV}.
\end{equation}

\section{Scalar mass eigenstates}\label{a:scalarSpectrum}
The scalar fields $\Phi_j$ ($j=1,2$), $\Phi_R$ can be parameterized in terms of gauge eigenstates as
\begin{align}
\Phi_j =& \begin{pmatrix} \dfrac{v_j+h^0_{j,11}+ia^0_{j,11}}{\sqrt2} & h^-_{j,12} \\ h^+_{j,21} & \dfrac{b_j+h^0_{j,22}+ia^0_{j,22}}{\sqrt2} \end{pmatrix}, \notag\\
\Phi_R =& \begin{pmatrix} h_R^+ & \dfrac{M+h^0_R+ia^0_R}{\sqrt2} \end{pmatrix}. \label{eq:phidef}
\end{align}
After spontaneous symmetry breaking, they are mixed into eigenstates of the mass matrix.

\subsection{Single-Bidoublet model}
In the SB model, only $\Phi_1$ and $\Phi_R$ are present, containing twelve real scalar components in total.
They are mixed into six Goldstone bosons and six massive particles.
The scalar masses are given in terms of the model parameters by
\begin{widetext}
\begin{align}
\frac{m_{h^0}^2}{v^2} =& \lambda_1 + \lambda_2c^2_{\beta_1} - \frac{(\lambda_4+\lambda_5s^2_{\beta_1})^2}{\lambda_3} + \ord{\xi^2},\qquad \frac{m_{H^0_1}^2}{M^2} = \frac{\lambda_5}{2c_{(2\beta_1)}} - \frac{\xi^2}{2} \Bigg( \lambda_2c^2_{(2\beta_1)} - \frac{\lambda_5^2s^2_{(2\beta_1)}c_{(2\beta_1)}}{\lambda_5-2\lambda_3c_{(2\beta_1})} + \ord{\xi^2} \Bigg), \notag\\
\frac{m_{H^0_2}^2}{M^2} =& \lambda_3 + \xi^2\Bigg( \frac{(\lambda_4+\lambda_5s^2_{\beta_1})^2}{\lambda_3} - \frac{\lambda_5^2s^2_{(2\beta_1)}c_{(2\beta_1)}}{\lambda_5-2\lambda_3c_{(2\beta_1)}} + \ord{\xi^2} \Bigg), \notag\\
\frac{m_{A^0}^2}{M^2} =& \frac{\lambda_5}{2c_{(2\beta_1)}} - \frac{\lambda_2}{2}\xi^2,\qquad \frac{m_{H^\pm}^2}{M^2} = \frac{\lambda_5}{2c_{(2\beta_1)}}\left( 1 + \xi^2c^2_{(2\beta_1)} \right). \label{eq:scalarmasses}
\end{align}
\end{widetext}
Here we define $h^0$ as the scalar that is the most $h^0_{1,11}$-like and $H^0_2$ as the scalar that is the most $h^0_R$-like.
The massive $CP$-odd state $A^0$ is a mixture of $a^0_{1,11}$ and $a^0_{1,22}$, and the charged states $H^\pm$ are a mixture of $h^\pm_{1,21}$ and $h^\pm_{1,12}$ with an $\ord{\xi}$ admixture of $h^\pm_R$.

\subsection{Complete LET model}
If we include $\Phi_2$ into the scalar sector, we have 20 real scalar components.
These are mixed into six Goldstone bosons, five massive $CP$-even states, three massive $CP$-odd states, and three pairs of massive charged scalars.
We define the $CP$-even mass eigenstates $h^0$, $H^0_1$, $H^0_2$, $H^0_3$, $H^0_4$ respectively as the most $h^0_{1,11}$-, $h^0_{2,11}$-, $h^0_{1,22}$-, $h^0_{2,22}$-, $h^0_R$-like scalars.
The massive $CP$-odd states $A^0_1$, $A^0_2$, $A^0_3$ are defined respectively as the most $a^0_{1,22}$-, $a^0_{2,11}$-, $a^0_R$-like scalars, and the charged states $H^\pm_1$, $H^\pm_2$, $H^\pm_3$ are defined respectively as the most $h^\pm_{2,21}$-, $h^\pm_{1,12}$-, $h^\pm_{2,12}$-like scalars.
Their masses are found to be
\begin{widetext}
\begin{align}
\frac{m_{h^0,H^0_1}^2}{v^2} =& \frac{1}{2}\Bigg( \Lambda_1c^2_\alpha + \Lambda_2s^2_\alpha \pm \sqrt{\left( \Lambda_1c^2_\alpha - \Lambda_2s^2_\alpha \right)^2 + \Lambda_3^2s^2_{(2\alpha)}} + \ord{\xi^2} \Bigg) \notag\\
\frac{m_{H^0_2}^2}{M^2} =& \frac{\lambda_5}{2c_{(2\beta_1)}} + \ord{\xi^2},\qquad \frac{m_{H^0_3}^2}{M^2} = \frac{\widetilde\lambda_4}{2c_{(2\beta_2)}} + \ord{\xi^2},\qquad \frac{m_{H^0_4}^2}{M^2} = \lambda_3 + \ord{\xi^2}, \notag\\
\frac{m_{A^0_1}^2}{v^2} =& -(\widetilde\lambda_7+\widetilde\lambda_{10})(c^2_{\beta_1}c^2_{\beta_2} + s^2_{\beta_1}s^2_{\beta_2}) - \frac{\widetilde\lambda_6}{2}s_{(2\beta_1)}s_{(2\beta_2)} + \ord{\xi^2}, \notag\\
\frac{m_{A^0_2}^2}{M^2} =& \frac{\lambda_5}{2c_{(2\beta_1)}} + \ord{\xi^2},\qquad \frac{m_{A^0_3}^2}{M^2} = \frac{\widetilde\lambda_4}{2c_{(2\beta_2)}} + \ord{\xi^2}, \notag\\
\frac{m_{H^\pm_1}^2}{v^2} =& -\frac12(\widetilde\lambda_6+\widetilde\lambda_7+\widetilde\lambda_{10})c^2_{(\beta_1-\beta_2)} - \frac{\widetilde\lambda_9}{2}c_{(2\beta_1)}c_{(2\beta_2)} + \ord{\xi^2}, \notag\\
\frac{m_{H^\pm_2}^2}{M^2} =& \frac{\lambda_5}{2c_{(2\beta_1)}} + \ord{\xi^2},\qquad \frac{m_{H^\pm_3}^2}{M^2} = \frac{\widetilde\lambda_4}{2c_{(2\beta_2)}} + \ord{\xi^2}. \label{eq:completeLETscalarMasses}
\end{align}
Here we have defined the parameter combinations
\begin{align}
\Lambda_1 \equiv& \lambda_1 + \lambda_2c^2_{\beta_1} - \frac{(\lambda_4+\lambda_5s^2_{\beta_1})^2}{\lambda_3},\qquad \Lambda_2 \equiv \widetilde\lambda_1 + \widetilde\lambda_2c^2_{\beta_2} - \frac{(\widetilde\lambda_3+\widetilde\lambda_4s^2_{\beta_2})^2}{\lambda_3}, \notag\\
\Lambda_3 \equiv& -\frac{(\lambda_4+\lambda_5s^2_{\beta_1})(\widetilde\lambda_3+\widetilde\lambda_4s^2_{\beta_2})}{\lambda_3} + \widetilde\lambda_5 + (\widetilde\lambda_6+\widetilde\lambda_7)c^2_{(\beta_1-\beta_2)} + (\widetilde\lambda_8+\widetilde\lambda_9+\widetilde\lambda_{10})(c^2_{\beta_1}c^2_{\beta_2} + s^2_{\beta_1}s^2_{\beta_2}).
\end{align}
\end{widetext}

\section{Higgs-coupling modifications of the SB model}\label{a:simplifiedLEThcms}
The SB model has a larger Higgs sector than the SM.
Since the SM-like scalar $h^0$ is a mixture of the various Higgs fields, its couplings generally differ from those of the SM.
These couplings depend on the scalar mixings, which in turn depend on the vevs and scalar-potential parameters.
We give the resulting Higgs-coupling modifications $\Delta_x$, as defined in \cref{eq:CMdefinition}, in the limit of small $\xi$.
The modifications of the tree-level couplings to $W$, $Z$, $t$, $b$ are given by
\begin{align}
\frac{\Delta_W}{\xi^2} =& -s^2_{(2\beta_1)} + \frac{\lambda_2s^2_{(4\beta_1)}}{8\lambda_5} + \frac{s_{(4\beta_1)}s_{(2\beta_1)}\lambda_{453}}{4} \notag\\
&- \frac{s^2_{\beta_1}\lambda_{453}^2}{2} + \ord{\xi},	\notag\\
\frac{\Delta_Z}{\xi^2} =& -c^4_{\theta_W^\prime} + \frac{\lambda_2s^2_{(4\beta_1)}}{8\lambda_5} + \frac{s_{(4\beta_1)}s_{(2\beta_1)}\lambda_{453}}{4} \notag\\
&- \frac{s^2_{\beta_1}\lambda_{453}^2}{2} + \ord{\xi},	\notag\\
\frac{\Delta_t}{\xi^2} =& -\frac{2\lambda_2s^2_{\beta_1} c^2_{(2\beta_1)}}{\lambda_5} - 2s^2_{\beta_1} c_{(2\beta_1)}\lambda_{453} - \frac{\lambda_{453}^2}{2} + \ord{\xi^2},	\notag\\
\frac{\Delta_b}{\xi^2} =& \frac{2\lambda_2c^2_{\beta_1} c^2_{(2\beta_1)}}{\lambda_5} + 2c^2_{\beta_1} c_{(2\beta_1)}\lambda_{453} - \frac{\lambda_{453}^2}{2} + \ord{\xi^2}.
\end{align}
Here we defined $\lambda_{453} = (\lambda_4+\lambda_5s^2_{\beta_1})/\lambda_3$.
The main contributions to the loop-induced photon coupling of the SM come from the $W$ and $t$ loops.
In the SB model, there is an additional contribution from the $H^\pm$ loop:
\begin{align}
\Delta_\gamma =& \frac{\xi^2A_0(\tau_{H^\pm})c_{(2\beta_1)}}{A_\text{SM}\lambda_5} \Bigg( \lambda_1 + \lambda_2(1+\frac12s^2_{(2\beta_1)}) + \lambda_5c_{(2\beta_1)} \notag\\
&\hspace{2cm}- \frac{\lambda_4(\lambda_4+\lambda_5c^2_{\beta_1})}{\lambda_3} + \ord{\xi^2} \Bigg).
\end{align}
Here $A_s(x)$ are the scalar loop functions, $\tau_x \equiv 4m_x^2/m_{h^0}^2$, and we defined the constant $A_\text{SM} \equiv A_1(\tau_W) + N_cQ_t^2A_{1/2}(\tau_t) = -6.5$.
The trilinear and quartic Higgs self-couplings are modified as well:
\begin{align}
\Delta_{\lambda_{3h}} =& \frac{-\lambda_2\lambda_3s^2_{\beta_1}c_{(2\beta_1)}}{\lambda_3(\lambda_1+\lambda_2c^2_{\beta_1}) - (\lambda_4+\lambda_5s^2_{\beta_1})^2} + \ord{\xi^2}, \notag\\
\Delta_{\lambda_{4h}} =& \frac{-\lambda_2\lambda_3s^2_{\beta_1}c_{(2\beta_1)} + (\lambda_4+\lambda_5s^2_{\beta_1})^2}{\lambda_3(\lambda_1+\lambda_2c^2_{\beta_1}) - (\lambda_4+\lambda_5s^2_{\beta_1})^2} + \ord{\xi^2}.
\end{align}
Note that the coupling modifications for vector bosons and fermions vanish in the small-$\xi$ limit, that is, the new physics decouples from the SM.
However, the modifications of the Higgs self-couplings only vanish for the region of parameter space where $\lambda_2 = 0$ and $\lambda_4 = -\lambda_5s^2_{\beta_1}$.
In general, the self-couplings are modified due to the more complicated structure of the scalar potential.

If we look for regions of parameter space where the coupling modifications become substantial, it is more useful to write the coupling modifications in terms of the scalar mixing coefficients $a_i$, as defined in \cref{eq:twinMixingCoefficients}:
\begin{align}
\Delta_{W,Z} =& c_{\beta_1}a_1 + s_{\beta_1}a_2 - 1 + \ord{\xi^2}, \notag\\
\Delta_t =& a_1/c_{\beta_1} - 1, \notag\\
\Delta_b =& a_2/s_{\beta_1} - 1.
\end{align}
The analogous coupling modifications $\Delta_x^\text{FP}$ of $H_2^0$ are obtained by substituting each $a_i$ by $a_i^\text{FP}$.
Since $\beta_1 = 0.0166$ is small, we can write $\Delta_W = \Delta_Z = \Delta_t \equiv \Delta_1$ and $\Delta_{W,Z}^\text{FP} = \Delta_t^\text{FP} \equiv \Delta_1^\text{FP}$, up to $\ord{\xi^2}$ and $\ord{\beta_1}$ corrections.

\bibliography{LETpaper}

\end{document}